\documentclass[pdflatex,sn-mathphys-num]{sn-jnl}
\usepackage{graphicx}%
\usepackage{multirow}%
\usepackage{amsmath,amssymb,amsfonts}%
\usepackage{amsthm}%
\usepackage{mathrsfs}%
\usepackage[title]{appendix}%
\usepackage{xcolor}%
\usepackage{textcomp}%
\usepackage{manyfoot}%
\usepackage{booktabs}%
\usepackage{algorithm}%
\usepackage{algorithmicx}%
\usepackage{algpseudocode}%
\usepackage{listings}%

\theoremstyle{thmstyleone}%

\theoremstyle{thmstyletwo}%

\theoremstyle{thmstylethree}%

\begin{document}
	
	\title[Article Title]{\bf Kantowski–Sachs Cosmological Model with Radiation and Generalized Chaplygin Gas}

	\author*[1]{\fnm{G.} \sur{Oliveira-Neto}}\email{gilnetojf@ufjf.br}
	
	\author[1]{\fnm{R.} \sur{Hincho Laucata}}\email{laucata.ruben@estudante.ufjf.br}

	
	\affil*[1]{\orgdiv{Departamento de F\'{\i}sica}, \orgname{Instituto de Ci\^{e}ncias Exatas}, \orgaddress{\street{Universidade Federal de Juiz de Fora}, \city{Juiz de Fora}, \postcode{CEP 36036-330}, \state{MG}, \country{Brazil}}}

	\abstract{In the present paper we study the evolution of a homogeneous and anisotropic Kantowski–Sachs cosmological model coupled with radiation and a generalized Chaplygin gas. Our main objective is verifying if this model evolves to a homogeneous and isotropic Friedmann-Robertson-Walker cosmological model, which is the more accepted model to describe our present Universe. If this is the case, we may imagine that the Kantowski–Sachs cosmological model could had been an early stage of our Universe.}

	
	
	\maketitle
	
	%%%%================================================================
    \section{Introduction}\label{Sec1}
	%%%%================================================================

    The most accepted cosmological model that describes the evolution of the Universe, since its beginning in the Big Bang, is the $\Lambda$-Cold Dark Matter model ($\Lambda$CDM). It gives a good account of several properties of our Universe. For instance, it explains the cosmic microwave background radiation, the large-scale distribution of matter, the observed abundance of light elements, and the accelerating expansion of the Universe. The $\Lambda$CDM model is based on three important postulates: the cosmological principle, Weyl's postulate, and general relativity (GR). The cosmological principle states that on large scales, the universe is homogeneous and isotropic, meaning that there is no special center or location. The Weyl's postulate says that the particles of the substratum lie in space-time on a congruence of timelike geodesics diverging from a point in the finite or infinite past. Finally, general relativity is the fundamental gravitational theory that describes the dynamics of the matter present in the model \cite{DinvernoVickers2022}.
    Many physicists believe today that the $\Lambda$CDM model is not sufficient to describe the very early universe, mainly because the Big Bang is a gravitational singularity
    that cannot be properly described by GR. When one turns to other gravitational theories to describe that epoch, like quantum gravity, one finds that the universe could have been very different from what it is today. In particular, the initial universe was not homogeneous and isotropic because at that moment the geometry of the Universe must have had a foam like structure, resonating between one configuration and another \cite{MTW1973}.
    
    Some physicists investigated the possibility that sometime after the Big Bang, when the dynamics of the matter present in the universe can already be described by GR, the cosmological principle is not yet valid. This means that the spacetime of cosmological models should not be described by the homogeneous and isotropic Friedmann-Robertson-Walker (FRW) metric. An initial and more simple possibility is to assume that the spacetime, at that time, is homogeneous but not isotropic. There are in the literature several different and well known homogeneous and anisotropic spacetimes, like the Bianchi spacetimes, the Kantowski-Sachs (KS) spacetime, etc. The most important condition that these early anisotropic models of the universe must satisfy is that they must go to the corresponding isotropic model after some time. That transformation must have occurred before the decoupling between matter and radiation, because the cosmic microwave background radiation, produced by the decoupling, is almost uniformly distributed in all directions with very small irregularities in different directions \cite{kester}.
    Some works in the literature on this subject investigated how different anisotropic cosmological models go to the corresponding isotropic FRW model \cite{barrow,adhav,simpliciano,gil0,parnovsky,thiago,thiago1,yuri}.

    Recent works in the literature study different aspects of anisotropic KS universes, at the classical and quantum levels.
    In Refs. \cite{faranas,faranas1}, the authors study KS spacetimes at the quantum level, using quantum cosmology. In the first work, the authors compute the {\it no-boundary} wavefunction of the model \cite{hawking} and in the second work they compute the {\it tunneling} wavefunction \cite{vilenking,vilenking1,vilenking2}. In both works, the authors use a path-integral approach. In the following works, the authors consider KS spacetimes, at the classical level.
    In Ref. \cite{gil0}, the authors investigate the dynamical evolution of a KS cosmological model coupled to a phantom perfect fluid and perfect fluids of radiation and dust. They show that, after some time, the model goes to a FRW cosmological model with positive spatial curvature.
    In Ref. \cite{singh}, the author considers a KS spacetime model coupled to stiff matter and a positive cosmological constant. After describing the evolution of the model, using dynamical system analysis, he finds that the model may evolve by expanding in an accelerated rate, by bouncing, or by re-collapsing.
    In Ref. \cite{paliathanasis}, the author study the dynamical evolution of a KS model coupled to a scalar field and a perfect fluid, using dynamical system analysis. He formulates the model in a Weyl integrable spacetime instead of a pseudo-Riemannian spacetime.
    In Ref. \cite{vinutha}, the authors investigate the dynamical and physical properties of a KS cosmological model coupled to a perfect fluid. The gravitational theory considered by the authors has an action in the functional form: $f(R,T) = R + \alpha R^2 + \lambda T$, where 
    $\alpha$ and $\lambda$ are constants, $R$ is the Ricci scalar, and $T$ is the trace of energy–momentum tensor. After imposing two conditions, they were able to obtain the solutions to the field equations of the model. With these solutions, they compute the dynamical and physical properties of the model.
    In Ref. \cite{oliveira}, the authors consider a KS cosmological model coupled to a radiation perfect fluid and a running cosmological constant (RCC). They explain that the RCC may be a consequence of the vacuum quantum effects of matter fields. They write the field equations of the model, including the energy-momentum conservation, and numerically solve them. After that, they investigated how the solutions depend on the RCC.
    In Ref. \cite{millano}, the authors study KS and Bianchi III cosmological models using a symmetric teleparallel $f(Q)$-theory, where $Q$ is the nonmetricity scalar. For a particular choice of $Q$, the authors investigated the asymptotic dynamics for the field equations, by means of a detailed analysis of the phase-space for the anisotropic cosmological models.
 
    As mentioned above, one well-known homogeneous and anisotropic spacetime is the Kantowski-Sachs (KS) one \cite{ks}. That spacetime has a spatial topology $S^2 \times \Re$ (or $S^2 \times S^1$, if the real line is compactified due to identifications), therefore, due to the spherical symmetry, one needs only two scale factors to describe it. That fact is one of the appealing properties of the KS spacetime. The curvatures of the spatial slices, of that spacetime, are constants and positives. Another important property of the KS spacetime is that it may describe the interior of a Schwarzchild black hole \cite{ks}.

	A very suitable way to write the KS metric is using the Misner parametrization \cite{misner}, which is given by,
	\begin{equation}
		ds^{2} = -dt^{2} + a(t)^{2} \left(e^{-\beta(t)} dr^{2} + e^{\beta(t)} (d\theta^{2} + \ \sin^{2}\theta \, d\phi^{2}) \right),
		\label{metric}
	\end{equation}
	where $a(t)$ is the scale factor and $\beta(t)$ is the anisotropic parameter, both depend only on time $t$, ($r, \theta, \phi$) are the spherical coordinates, and we are using the natural unit system, where $c=8\pi G=1$. The Misner parametrization is very useful when one is studying the isotropization of the model. Observing Eq. (\ref{metric}), it is not difficult to notice that when $\beta(t)$ goes to nil or a constant, the metric goes to the FRW metric with constant positively curved spatial sections. In these limits $a(t)$ Eq. (\ref{metric}) goes to the FRW scale factor and $\beta(t)$ Eq. (\ref{metric}) gives a measure of the KS anisotropy.

    An important discovery took place in 1998 after astronomers observed distant supernovae: the Universe is expanding in an accelerated rate \cite{riess0,perlmutter}. This discovery started an intense debate about the cause of that acceleration. If the universe is a closed system and gravity is the fundamental interaction acting on matter and radiation in the universe, what could be causing such an acceleration? A major line of research considers that the cause of that accelerated expansion is due to the presence of a new type of matter in the universe. That new type of matter must have properties very different from ordinary matter. One such property is that it must be repulsive rather than attractive with respect to the gravitational interaction. Due to that strange property some people called it `exotic' matter or dark energy. Two interesting examples of dark energy are the Chaplygin gas (CG) \cite{chaplygin,kamenshchik2001alternative} and its generalization the generalized Chaplygin gas (GCG) \cite{kamenshchik2001alternative,bilic,bento}. The energy density of the GCG depends on three parameters $A$, $C$ and $\alpha$. When one sets $\alpha=1$, one recovers the CG. 
    In the literature on classical and quantum cosmology, several important works have been produced studying both gases CG and GCG \cite{paulo,gil,pedram0,pedram,majumder,pedram1,pedram2,monerat}.

    In the present work, we study the dynamical evolution of an initially homogenous and anisotropic KS model in the presence of a generalized Chaplygin gas and a perfect fluid of radiation. The perfect fluid of radiation was included in our model because it is believed that it was very important in the initial moments of the universe. We start writing the Lagrangian and then the Hamiltonian of the model. From the Hamiltonian, we obtain the Hamilton's equations of the model. They are equivalent to Einstein's equations. Before solving Einstein's equations, we use the Hamiltonian to draw phase portraits of the model. They help us to obtain the qualitative behaviors of the solutions to the scale factor $a(t)$ and the  anisotropic parameter $\beta(t)$. Then, we solve Einstein's equations numerically and study how each parameter modifies the evolution of the scale factor $a(t)$ and the  anisotropic parameter $\beta(t)$. In particular, we find that $a(t)$ expands in an accelerated rate and $\beta(t)$ becomes a constant after some time. This means that the present homogeneous and anisotropic KS model goes to a homogeneous and isotropic FRW model after some time.

    In Section \ref{Sec2}, we compute the Lagrangian and Hamiltonian of the KS model coupled to a generalized Chaplygin gas and radiation. Then, using that Hamiltonian, we derive the coupled system of differential equations for the metric variables. In Section \ref{Sec3}, we build phase portraits of the model, to give a general idea of the dynamical behaviors of the metric functions. In Section \ref{Sec4}, we numerically solve the coupled differential equation system and determine how the scale factor $a(t)$ and the anisotropic parameter $\beta(t)$ behave, as functions of the time coordinate and the parameters $\alpha$, $A$, $C$, $C_r$.
    %and the initial conditions $a(t=0) \equiv a_0$, $\dot{a}(t=0) \equiv \dot{a}_0$ and $\beta(t=0) \equiv \beta_0$. 
    We pay special attention to the solution that represents the expansion. In Section \ref{Sec5}, we present the dynamical evolution of the model using the most accepted present values of the parameters $A$, $C$, $\alpha$ and $C_r$.
    Finally, in Section \ref{Sec6} we give our conclusions.

    %%%%%%%%%%%%%%%%%%%%%%%%%%%%%%%%%%%%%%%%%%%%%%%%%%%%%%%%%%%%%
	\section{The Hamiltonian and system of differential equations of the model} \label{Sec2}
    %%%%%%%%%%%%%%%%%%%%%%%%%%%%%%%%%%%%%%%%%%%%%%%%%%%%%%%%%%%%%
    
	Since our interest is in investigating the homogeneous and anisotropic Kantowski–Sachs ($KS$) cosmological model, whose metric is expressed by Eq.~\eqref{metric}, we consider that the matter content is described by a generalized Chaplygin gas and a perfect radiation fluid. The four-velocity of the fluid and the gas is given by $U^\mu = (1, 0, 0, 0)$, which indicates that they are at rest in the adopted comoving coordinate system. This implies the absence of heat flows and viscous stresses, that is, the effects of internal interaction are neglected \cite{Wald1984}. Under these conditions, the energy-momentum tensor $T_{\mu\nu}$ takes the form:
	\begin{equation}
		T_{\mu\nu}=(\rho+p)U_\mu U_\nu+pg_{\mu\nu},
		\label{Tensor de energia momento}
	\end{equation}
	where $\rho$ and $p$ represent, respectively, the energy density and pressure of the fluid and gas. In this case, the following equation of state is assumed for the fluid:
	\begin{equation}
		p_r=\omega \rho_r.
		\label{Eq. de estado}
	\end{equation}
	Here $\omega$, is the parameter that characterizes the type of perfect fluid, $\rho_r$ and $p_r$ represent, respectively, the energy density and pressure of the fluid. The material content in this work is composed of a radiation perfect fluid ($\omega = 1/3$), which is believed to have dominated the initial instants of the Universe \cite{Weinberg2008} and a generalized Chaplygin gas. The GCG emerges as a natural candidate for the unification of dark matter and dark energy, the latter being associated with the accelerated expansion of the Universe in late cosmological epochs \cite{bento}. The equation of state of this gas is defined as:  
	\begin{equation}
		p_c = -\frac{A}{\rho_c^\alpha},
		\label{Chaplygin}
	\end{equation}
	where $p_c$ and $\rho_c$ represent, respectively, the pressure and energy density of the generalized Chaplygin gas. The parameter $A$ is a positive constant, and $\alpha$ is a dimensionless parameter in the range $0<\alpha \leq 1$. In the particular case where $\alpha=1$, the case of the Chaplygin gas \cite{kamenshchik2001alternative} is recovered.
	
	To investigate the dynamics of the model, we start from the formulation of Einstein's equations using the Hamiltonian formalism. The starting point is the Einstein-Hilbert action, represented as $(S)$.
	\begin{equation}
		S=\int\left[\frac{R}{2}+\mathcal{L}_{m}\right]\sqrt{-g}dx^4,
		\label{Acao}
	\end{equation}
	where $R$ is the curvature scalar, $\mathcal{L}_{m}$ is the Lagrangian density associated with matter, and $g$ represents the determinant of the metric Eq.\eqref{metric}. From the metric, the curvature scalar can be written as:
	\begin{equation}
		R= 6\frac{\ddot{a}}{a}+\ddot{\beta}+4\frac{\dot{a}\dot{\beta}}{a}+\dot{\beta}^2+6\frac{\dot{a}^2}{a^2}+2\frac{e^{-\beta}}{a^2}.
		\label{Escalar de Ricci}
	\end{equation}
	For the present model, we assume that $\mathcal{L}_{m}$ is,
	\begin{equation}
		\mathcal{L}_{m}=-\rho.
		\label{Densidade lagrangiana da materia}
	\end{equation}
	Where $\rho$ is the total energy density. To write $\rho$ in terms of $a(t)$ and $\beta(t)$, we have to write the conservation equation for the energy-momentum tensor Eq.\eqref{Tensor de energia momento}. This equation tells us that the divergence of this tensor is zero \cite{DinvernoVickers2022},
    \begin{equation}
        \nabla_\nu T^{\mu\nu}=0.
        \label{conservation}
    \end{equation}
    A necessary condition for writing this conservation equation is knowing the components of the energy-momentum tensor. Thus, using the metric $KS$ Eq. \eqref{metric} and the definition of the four-velocity $U^\mu=(1,0,0,0)$, we obtain the non-zero components of the tensor Eq.\eqref{Tensor de energia momento}:
    \begin{equation}
        T^{tt} = \rho,\; T^{rr} = pa^{-2}e^{\beta},\; T^{\theta\theta} = pa^{-2}e^{-\beta},\; T^{\phi\phi} = pa^{-2}e^{-\beta}\ \sin^{-2}(\theta).
        \label{Os tensores de momento do K-S}
    \end{equation}
	For simplicity, let us consider that the pressures of the radiation fluid and the generalized Chaplygin gas are isotropic, so the pressure $p$ is the same in all directions. Now, from the conservation of the energy-momentum tensor Eq. \eqref{conservation}, with the help of the non-zero components of $T_{\mu\nu}$ Eq. \eqref{Os tensores de momento do K-S}, we obtain the total energy density $\rho(t)$ of the model in terms of $a(t)$ and $\beta(t)$. Taking into account Eqs.\eqref{Eq. de estado} and \eqref{Chaplygin}, $\rho(t)$ is given by:
	\begin{equation}
		\rho=C_r\frac{e^{-2\beta/3}}{a^4}+\bigg( A+C(a^3e^{\beta/2})^{-(\alpha+1)}\bigg)^{\frac{1}{\alpha+1}},
		\label{Densidade total}
	\end{equation}
	where $C_r$ is an integration constant associated with the radiation energy density, while $C$ is an integration constant associated with the generalized Chaplygin gas. Then, substituting Eqs. \eqref{Escalar de Ricci}, \eqref{Densidade lagrangiana da materia} and \eqref{Densidade total} into the action Eq. \eqref{Acao}, we find:
	\begin{equation}
		S=\int\mathcal{L}_{eff} dt.
		\label{Acao Reducido}
	\end{equation}
	Where $\mathcal{L}_{eff}$ is the effective Lagrangian density, which represents the sum of the gravitational and matter contributions. After an integration by parts, $\mathcal{L}_{eff}$ can be rewritten as:
	\begin{equation}
		\begin{aligned}
			\mathcal{L}_{eff}=\frac{ae^{\beta/2}}{2}\bigg[(\frac{1}{2}a^2\dot{\beta}^2-6\dot{a}^2-2a\dot{a}\dot{\beta}+2e^{-\beta})-2C_r~a^{-2}e^{-2\beta/3}\\
			-2\bigg( A+C(a^3e^{\beta/2})^{-(\alpha+1)}\bigg)^{\frac{1}{\alpha+1}}a^2 \bigg].
		\end{aligned}
		\label{Lagrangiano effetivo}
	\end{equation}
	Now, we want to determine the Hamiltonian of the model. Therefore, we calculate the canonical momenta conjugate to the factors $a(t)$ and $\beta(t)$, from Eq. ~\eqref{Lagrangiano effetivo}:
	\begin{equation}
		\begin{aligned}
			P_a&=\frac{\partial \mathcal{L}_{eff}}{\dot{a}}=\frac{1}{2}ae^{\beta/2}(-12\dot{a}-2a\dot{\beta}),\\
			P_{\beta}&=\frac{\partial\mathcal{L}_{eff}}{\dot{\beta}}=\frac{1}{2}a^2e^{\beta/2}(a\dot{\beta}-2\dot{a}).
		\end{aligned}
		\label{Momentos conjugados do L_eff}
	\end{equation}
	By inverting equations \eqref{Momentos conjugados do L_eff}, we can determine the `velocities' $\dot{a}(t)$ and $\dot{\beta}(t)$ as functions of the conjugated momenta:
	\begin{equation}
		\begin{aligned}
			\dot{a}&=-\frac{e^{-\beta/2}}{8a^2}(aP_a+2P_{\beta}),\\
			\dot{\beta}&=\frac{e^{-\beta/2}}{4a^3}(6P_{\beta}-aP_a).
		\end{aligned}
		\label{Os fatores a(t), beta(t)}
	\end{equation}
	The effective Hamiltonian of the model ($\mathcal{H}_{eff}$) can be obtained by using the standard definition:
	\begin{equation}
		\mathcal{H}_{eff}=P_a\dot{a}+P_\beta\dot{\beta}-\mathcal{L}_{eff}.
		\label{Formula do H_eff}
	\end{equation}
	Now, by substituting Eqs.\eqref{Lagrangiano effetivo} and \eqref{Os fatores a(t), beta(t)} into Eq.\eqref{Formula do H_eff}, we explicitly obtain the effective Hamiltonian of the model:
	\begin{equation}
		\begin{aligned}
			\mathcal{H}_{eff}=e^{-\beta/2}\bigg(\frac{3P_{\beta}^2}{4a^3}-\frac{P_a^2}{16a}-\frac{P_aP_{\beta}}{4a^2}-a \bigg)+C_r\frac{e^{-\beta/6}}{a}\\ +\bigg( A+C(a^3e^{\beta/2})^{-(\alpha+1)}\bigg)^{\frac{1}{\alpha+1}}a^3e^{\beta/2}.
		\end{aligned}
		\label{Hamiltoniano efetivo de K-S}
	\end{equation}
	General relativity is a theory with constraints \cite{Sundermeyer1982}. This means that in addition to the dynamical equations, it must satisfy other equations due to these constraints. In the present model, the only constraint
    is given by:
	\begin{equation}
		\mathcal{H}_{eff}=0.
		\label{1 retricao}
	\end{equation}
	This equation is known in the literature as the super-Hamiltonian constraint \cite{MTW1973}. In this context, it is equivalent to the temporal component of Einstein's equations:  $G_{00}=T_{00}$, where $G_{\mu \nu}$ is the Einstein's tensor and the index $0$ represents the time component. By imposing the constraint Eq.\eqref{1 retricao}, rewriting the canonical moments in terms of $\dot{a}$ and $\dot{\beta}$ with the aid of Eqs. \eqref{Momentos conjugados do L_eff}, and multiplying the result by $(-1)e^{\beta/2}a^{-3}$, we obtain the equation:
	\begin{equation}
		\frac{-\dot{\beta}^2}{4}+\frac{\dot{a}\dot{\beta}}{a}+\frac{3\dot{a}^2}{a^2}+\frac{e^{-\beta}}{a^2}=C_r\frac{e^{-2\beta/3}}{a^4}+\Bigg[A+C\bigg(a^3e^{\beta/2}\bigg)^{-\alpha-1}\Bigg]^{\frac{1}{\alpha+1}}.
		\label{G_00}
	\end{equation}
	Starting from the effective Hamiltonian Eq.\eqref{Hamiltoniano efetivo de K-S}, we derive Hamilton's equations to obtain the other independent field equations. In this context, the equations of motion are written as follows:
	\begin{equation}
		\dot{a}=\frac{\partial \mathcal{H}_{eff}}{\partial P_a}=-\frac{e^{-\beta/2}}{8a^2}(aP_a+2P_{\beta}),
		\label{1 EQ. de Movimento}
	\end{equation}
	\begin{equation}
		\begin{aligned}
			\dot{P}_{a}&=e^{-\beta/2} \bigg( \frac{9P_{\beta}^2}{4a^4} - \frac{P_a^2}{16a^2} - \frac{P_a P_{\beta}}{2a^3} +1 \bigg) + \frac{C_r e^{-\beta/6}}{a^2} - 3a^2e^{\beta/2} \bigg( A + C (a^3 e^{\beta/2})^{-(\alpha+1)} \bigg)^{\frac{1}{\alpha+1}}\\&+ 3Ca^2 e^{\beta/2} (a^3 e^{\beta/2})^{-(\alpha+1)} \bigg( A + C (a^3 e^{\beta/2})^{-(\alpha+1)} \bigg)^{\frac{-\alpha}{\alpha+1}},
		\end{aligned}
		\label{2 EQ. de Movimento}
	\end{equation}
	\begin{equation}
		\begin{aligned}
			\dot{\beta}&=\frac{\partial \mathcal{H}_{eff}}{\partial P_{\beta}}=\frac{e^{-\beta/2}}{4a^3}(6P_{\beta}-aP_a),
		\end{aligned}
		\label{3 EQ. de Movimento}
	\end{equation}
	\begin{equation}
		\begin{aligned}
			\dot{P}_{\beta}&= \frac{e^{-\beta/2}}{2} \bigg( \frac{3P_{\beta}^2}{4a^3} - \frac{P_a^2}{16a} - \frac{P_a P_{\beta}}{4a^2} - a \bigg)
			+ \frac{C_r e^{-\beta/6}}{6a} - \frac{a^3}{2}e^{\beta/2} \bigg( A + C (a^3 e^{\beta/2})^{-(\alpha+1)} \bigg)^{\frac{1}{\alpha+1}}\\& + \frac{C}{2} a^3 e^{\beta/2} (a^3 e^{\beta/2})^{-(\alpha+1)} 
			\bigg( A + C (a^3 e^{\beta/2})^{-(\alpha+1)} \bigg)^{\frac{-\alpha}{\alpha+1}}.
		\end{aligned}
		\label{4 EQ. de Movimento}
	\end{equation}
	Based on these expressions, we can obtain the independent field equations. First, by calculating the second derivative of Eq. \eqref{1 EQ. de Movimento}. Then, substituting the values of the first time derivatives of the canonical moments  $\dot{P_a}$ and $\dot{P_\beta}$ from Eqs. \eqref{2 EQ. de Movimento} and \eqref{4 EQ. de Movimento}, respectively, and multiplying the result by $1/(4a)$, we obtain:
	\begin{equation}
		-2a\ddot{a}-\dot{a}^2-\frac{a^{2}\dot{\beta}^2}{4}=C_r\frac{e^{-2\beta/3}}{3a^2}-a^2A\bigg[A+C\bigg(a^3e^{\beta/2}  \bigg)^{-\alpha-1}\bigg]^{(\frac{-\alpha}{\alpha+1})}.
		\label{1eq Importante}
	\end{equation}
	Similarly, we calculate the second derivative of Eq. \eqref{3 EQ. de Movimento}, and substitute the values of $\dot{P_a}$ and $\dot{P}_\beta$ from Eqs. \eqref{2 EQ. de Movimento} and \eqref{4 EQ. de Movimento}. Finally, multiplying the result by $1/(4a)$, we obtain:
	\begin{equation}
		3a\dot{a}\dot{\beta}+a^2\ddot{\beta}=-\frac{a^2\dot{\beta}^2}{2}-e^{-\beta}.
		\label{2eq Importante}
	\end{equation}
	Adding Eqs.\eqref{1eq Importante} and \eqref{2eq Importante}, and rewriting the right side of the resulting equation in terms of total pressure $p=p_c+p_r$, where $p_c$ and $p_r$ are the generalized Chaplygin gas and radiation pressures, respectively. In this way, we obtain the following equation:
	\begin{equation}
		a^2\ddot{\beta}+2a\ddot{a}+\dot{a}^2+\frac{3a^2\dot{\beta}^2}{4}+3a\dot{a}\dot{\beta}+e^{-\beta}=-a^2p.
		\label{3eq Importante}
	\end{equation}
	Using Eqs. \eqref{Eq. de estado}, \eqref{Chaplygin} and \eqref{Densidade total}, we can explicitly rewrite $p$ as:
	\begin{equation}
		p=C_r\frac{e^{-2\beta/3}}{3a^4}-A\bigg[A+C\bigg(a^3e^{\beta/2}  \bigg)^{-\alpha-1}\bigg]^{(\frac{-\alpha}{\alpha+1})}.
		\label{Pessao}
	\end{equation}
	Eq.\eqref{3eq Importante} corresponds to the $(r,r)$ component of Einstein's field equations. The last independent equation is obtained by rewriting the right-hand side of Eq.\eqref{1eq Importante} in terms of pressure $p$, as given in Eq.\eqref{Pessao}, resulting in the following equation:
	\begin{equation}
		2a\ddot{a}+\dot{a}^2+\frac{a^2\dot{\beta}^2}{4}=-a^2p.
		\label{4eq Importante}
	\end{equation}
	The equation \eqref{4eq Importante} corresponds to the components $(\theta,\theta),$ and $(\phi,\phi)$ of Einstein's field equations, which are equivalent in this model due to spherical symmetry in the angular component.
	
	Equations \eqref{G_00}, \eqref{3eq Importante}, and \eqref{4eq Importante} form a system of second-order ordinary differential equations. To solve it, one must provide initial conditions for $a(t) (a_0), \beta(t) (\beta_0), \dot{a}(t) (\dot{a}_0)$ and $\dot{\beta}(t) (\dot{\beta}_0)$. However, it is observed that there are three equations for only two variables $a(t),\beta(t)$. Thus, we only need two equations to describe the evolution of the system. Therefore, we choose Equation \eqref{G_00} (time component), along with the following linear combination between Eqs. \eqref{3eq Importante} and \eqref{4eq Importante}:
	\begin{equation}
		a^2\ddot{\beta}+\frac{a^2\dot{\beta}^2}{2}+3a\dot{a}\dot{\beta}+e^{-\beta}=0.
		\label{5Eq importante}
	\end{equation}
	Using the first-order equation, Eq.\eqref{G_00}, we determine the physically consistent initial conditions. By fixing the values of the initial conditions $a_0, \beta_0, \dot{a_0}$ and the parameters $C, A, C_r$ and $\alpha$, we obtain from Eq.\eqref{G_00}, the value of the initial condition $\dot{\beta_0}$ that physically satisfies the system.
    
    %%%%%%%%%%%%%%%%%%%%%%%%%%%%%%%%%%%%%%%%%%%%%%%%%%%%%%%%%%%%%%%%
	\section{Qualitative study of the solutions}\label{Sec3}
	%%%%%%%%%%%%%%%%%%%%%%%%%%%%%%%%%%%%%%%%%%%%%%%%%%%%%%%%%%%%%%%%
	In this section, we qualitatively study the solutions to the present model. We do that by constructing phase portraits in the planes $(a,P_a)$ and $(\beta,P_ \beta)$, associated with the present model. We start by imposing the constraint Eq. (\ref{1 retricao}), with the aid of Eq. (\ref{Hamiltoniano efetivo de K-S}). After solving Eq. (\ref{1 retricao}) for several different values of the parameters $C$, $A$, $C_r$, $\alpha$ and the initial conditions $a_0$, $\dot{a}_0$, $\beta_0$ e $\dot{\beta}_0$, we notice that the solutions are expansive. More precisely, the solutions behave as follows. When the scaling factor $a(t)$ assumes small values, the Universe expands rapidly from an initial singularity at $a(t)\to 0$. Subsequently, the expansion rate tends to smooth out as a result of the phase transition process between the radiation fluid and the generalized Chaplygin gas. In this regime, gravitational attraction ceases to dominate, while the Chaplygin gas begins to exert a repulsive effect, leading to a new regime of accelerated expansion. An example of this behavior is shown in Fig~\ref{D_fases (a,P)}. On the other hand, the anisotropy parameter $\beta(t)$ grows rapidly from $\beta(t)=0$, until it asymptotically reaches a constant positive value. This regime indicates that the model naturally tends towards isotropization, as illustrated in  Fig~\ref{D_fases (beta,P)}. Since our main interest lies in the behavior of solutions for larger values of the scale factor $a(t)$, where the isotropization process effectively occurs, we will not consider the behavior of solutions corresponding to very small values when $a(t)\to 0$, close to the initial singularity.
    
	The solutions obtained are shown in Figs. \ref{D_fases (a,P)} and \ref{D_fases (beta,P)}. To represent these figures, we considered different values of the parameter $C_r$ associated with the radiation fluid, keeping the parameters $\alpha=0.7$ and $C=A=1$ fixed. The results show that the Universe emerges from an initial singularity, undergoes a phase transition, and then the scale factor $a(t)$ begins an accelerated expansion, while the anisotropy parameter $\beta(t)$ tends to a constant value. This behavior shows that the model naturally evolves towards a homogeneous and isotropic state. In the following sections, we will analyze the numerical results in detail, with the aim of verifying and corroborating the qualitative conclusions discussed here, systematically investigating the influence of the parameters on the dynamic evolution of the model.
	
	\begin{figure}[h]
		\centering
		\includegraphics[width=0.5\textwidth]{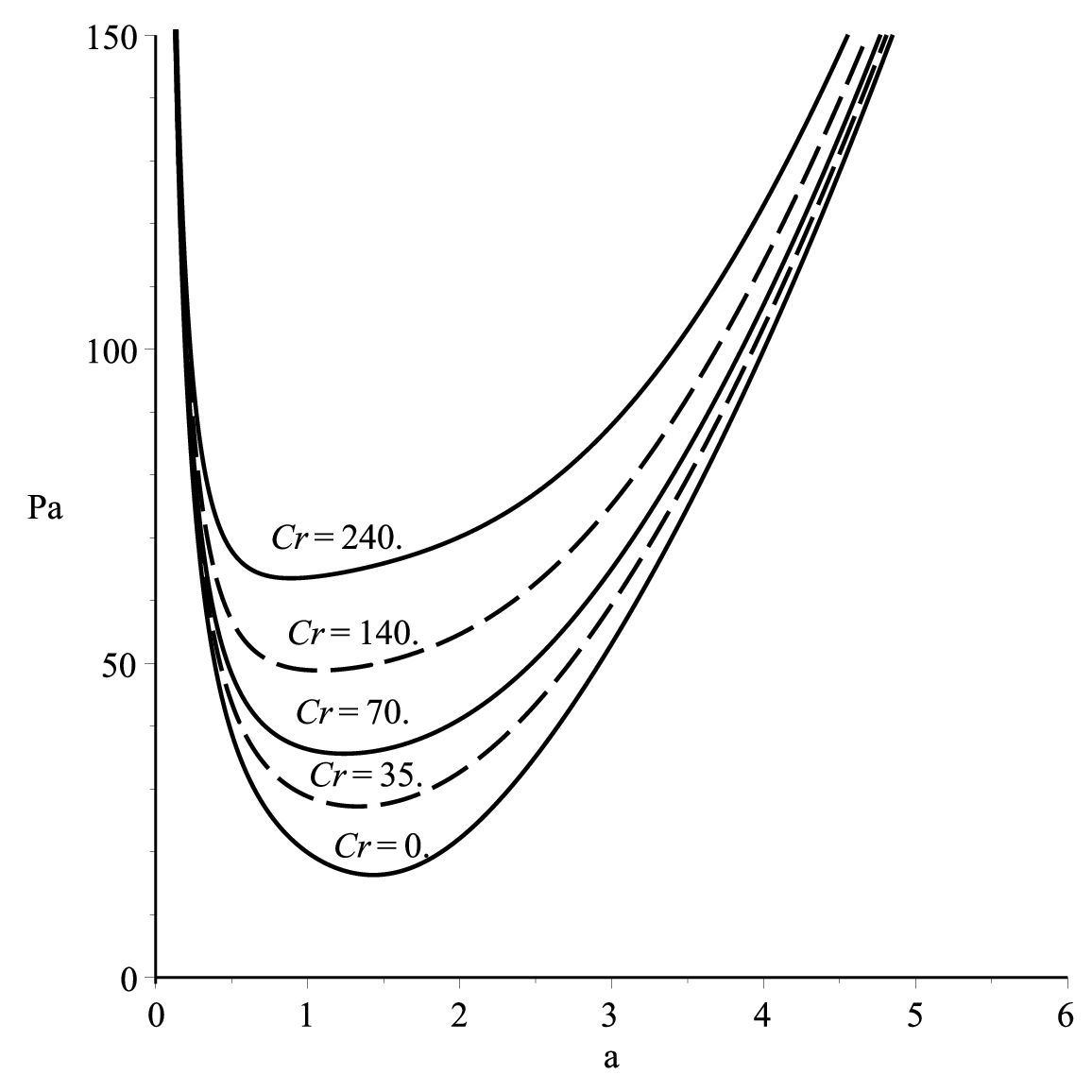}
		\caption{Phase-space representation $(a,Pa)$ for different values of $C_r=0,35,70,140,240$ with $A=C=1$, $\alpha=0.7$, $\beta_{0}=1$ and $P_\beta=9.70109722$. }
		\label{D_fases (a,P)}
	\end{figure}
	\begin{figure}[h]
		\centering
		\includegraphics[width=0.5\textwidth]{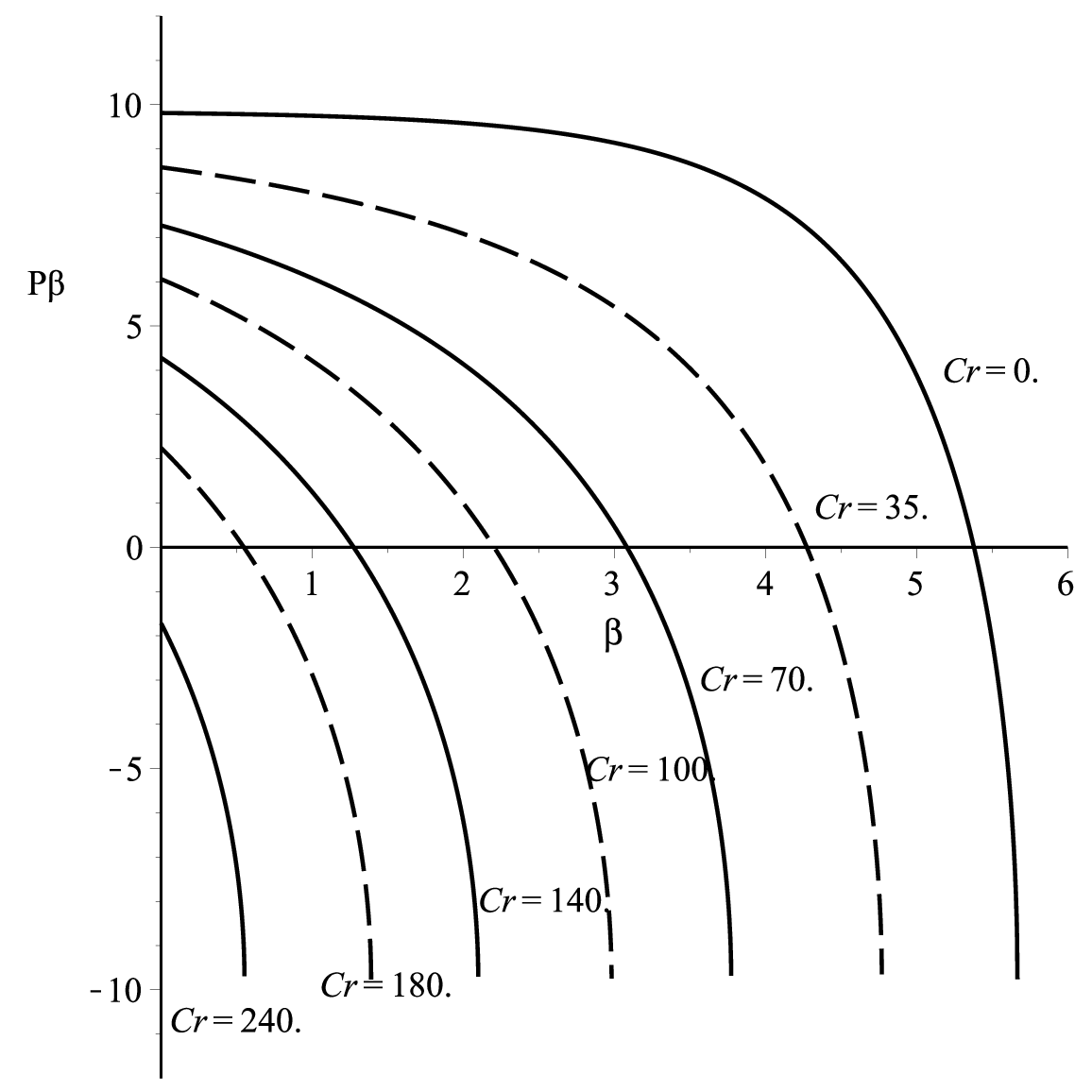}
		\caption{Phase-space representation $(\beta,P_\beta)$ for different values of $C_r=0,35,70,140,240$ with $A=C=1$, $\alpha=0.7$, $a_{0}=1$ and $P_a=-58.9715049$. }
		\label{D_fases (beta,P)}
	\end{figure}
    
	%%%%%%%%%%%%%%%%%%%%%%%%%%%%%%%%%%%%%%%%%%%%%%%%%%%%%%%%%%%%%%%%
	\section{Quantitative analysis of the numerical solutions of the model}\label{Sec4}
    %%%%%%%%%%%%%%%%%%%%%%%%%%%%%%%%%%%%%%%%%%%%%%%%%%%%%%%%%%%%%%%%
	In this section, we solve numerically the system of second order ordinary differential equations formed by Eqs.~\eqref{5Eq importante} and \eqref{G_00}. We solve the system of equations for several different values of the parameters $C$, $A$, $C_r$, $\alpha$. 
    We want to investigate how the variation of these parameters  quantitatively modifies the dynamic evolution of $a(t)$ and $\beta(t)$. To do this, we varied the value of one of the parameters and kept the values of all other quantities fixed. In all examples presented in the following subsections, the values of the parameters were selected to provide a better visualization of the results. In particular, we adopted the following reference values for the initial conditions
	\begin{equation}
		a_0=1,\qquad  \dot{a}_0=1,\qquad \beta_0=1
		\label{CONDICOES INICIAIS}
	\end{equation}
	
	Physically consistent initial conditions are determined in each case by leaving the initial rate of change of the anisotropy parameter $\dot{\beta}_0$ as a free quantity, which is subsequently calculated from the first-order differential equation given by Eq.~\eqref{G_00}. As we shall see next, the quantitative behavior of the solutions agrees with the qualitative results we obtained in the previous Section \ref{Sec3}.
	
	%%%%%%%%%%%%%%%%%%%%%%%%%%%%%%%%%%%%%%%%%%%%%%%%%%%%%
	%%                   Resultados                    %%
	%%%%%%%%%%%%%%%%%%%%%%%%%%%%%%%%%%%%%%%%%%%%%%%%%%%%%
	
    %%%%%%%%%%%%%%%%%%%%%%%%%%%%%%%%%%%%%%%%%%%%%
    \subsection{Studying the influence of $\alpha$}\label{Sec4.1}
    %%%%%%%%%%%%%%%%%%%%%%%%%%%%%%%%%%%%%%%%%%%%%
    
\begin{figure}[h!]
    	\centering
    	\includegraphics[width=0.7\textwidth]{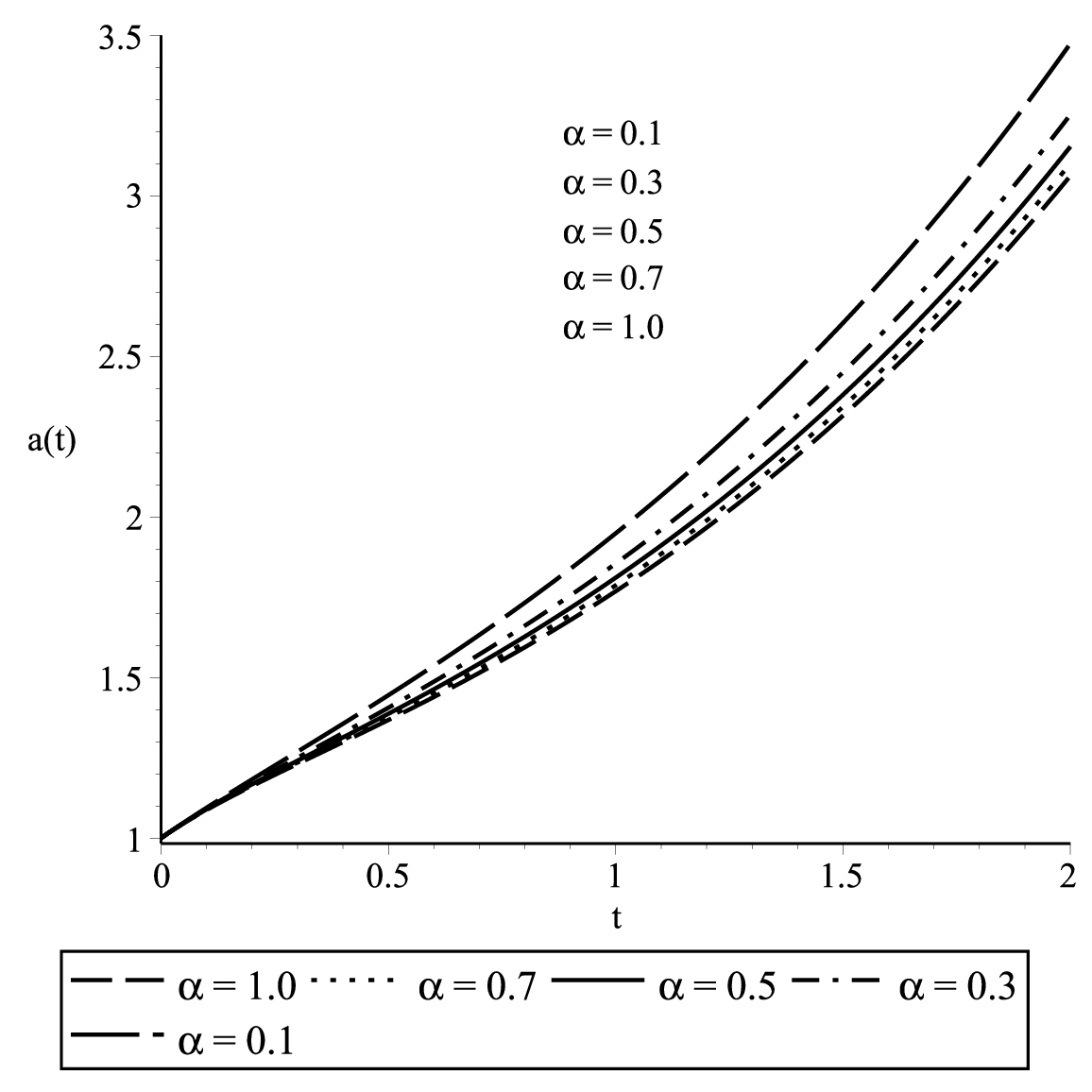}
    	\caption{$a(t)$ as a function of $t$ for different values of $\alpha$ with $A = C_r = 1, C = 5, \alpha = 0.7$ and $a_0 = \beta_0 = \dot{a}_0 = 1$.}
    	\label{fig_a. Variando alpha}
    \end{figure}

\begin{figure}[h!]
    	\centering
    	\includegraphics[width=0.7\textwidth]{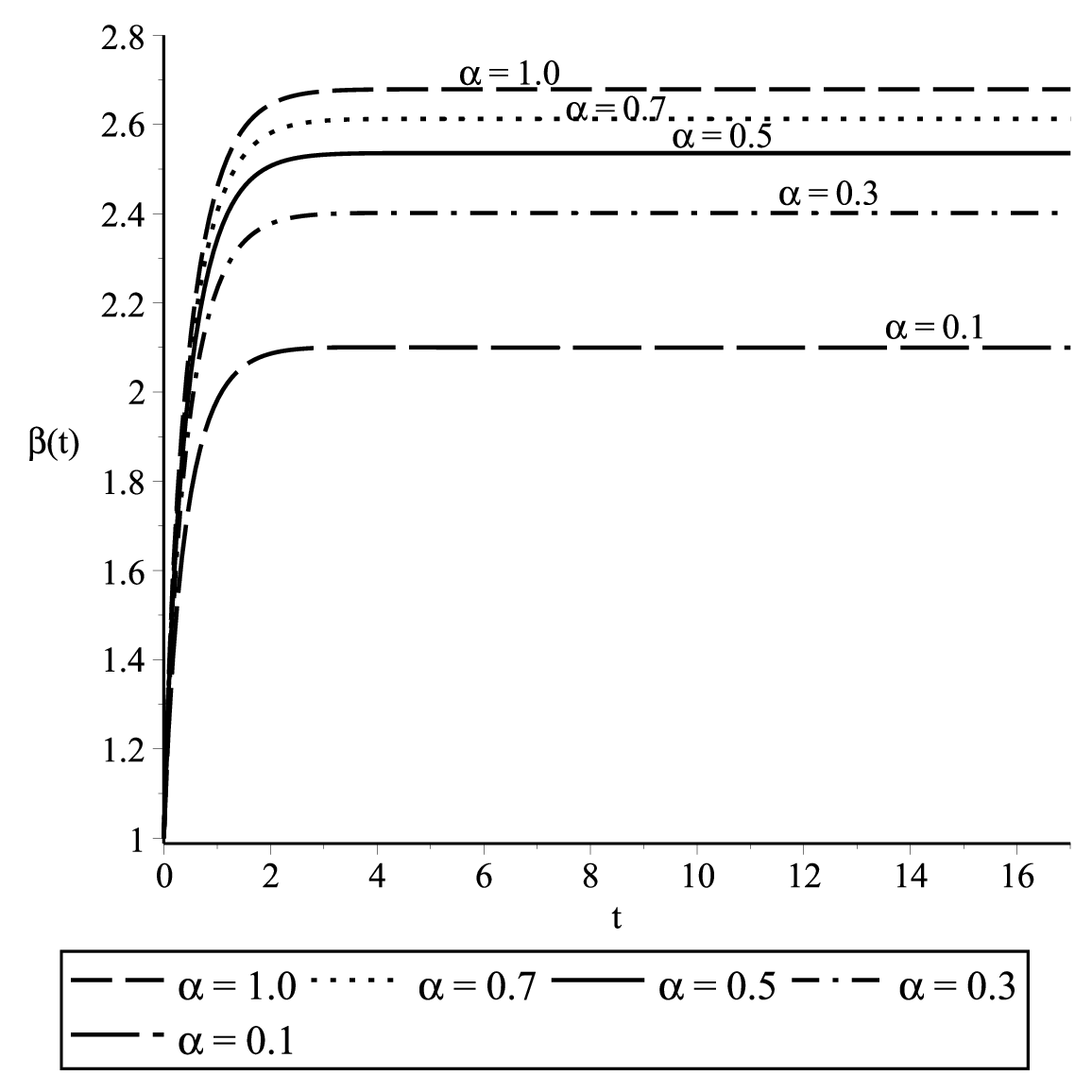}
    	\caption{$\beta(t)$ as a function of $t$ for different values of $\alpha$ with $A=C_r=1,C=5, \alpha=0.7$ and $a_0=\beta_0=\dot{a}_0=1$.}
    	\label{fig_b. Variando alpha}
    \end{figure}

    We begin by analyzing how the parameter $\alpha$ of the generalized Chaplygin gas influences the dynamic evolution of $a(t)$ and $\beta(t)$, after solving the system of equations \eqref{G_00} and \eqref{5Eq importante}. Changing the value of $\alpha$, while keeping the parameters $C$, $A$, and $C_r$ fixed, we find that the scale factor $a(t)$ expands more rapidly for smaller values of $\alpha$, as illustrated in Fig.~\ref{fig_a. Variando alpha}. This behavior is expected because, in this regime, the generalized Chaplygin gas becomes increasingly gravitationally repulsive. On the other hand, the anisotropy parameter $\beta(t)$ converges asymptotically to a positive constant value; the value depends on the parameter $\alpha$. For smaller values of $\alpha$, $\beta(t)$ asymptotically tends towards smaller values; while for larger values, $\beta(t)$ stabilizes at relatively larger values, as illustrated in Fig.~\ref{fig_b. Variando alpha}. The behavior of the system is corroborated by the data presented in Table~\ref{Tabela do parametro alpha}. 
    In this Table, we introduce the time $t_s$ for the first time. The time $t_s$ is numerically determined as the instant when the anisotropy parameter $\beta(t)$ enters an asymptotic regime. More precisely, we define $t_s$ as the time when the rate of change of the anisotropic parameter becomes practically zero, that is, $\dot{\beta}(t_s) \approx 0$, assuming very small values on the order of $10^{-12}$ or smaller. From this instant onward, the anisotropy parameter $\beta(t)$ remains constant, and the Kantowski–Sachs metric effectively reduces to the Friedmann–Robertson–Walker metric with positive spatial curvature ($k = +1$), characterizing the beginning of the isotropic regime of the Universe.
    In Table~\ref{Tabela do parametro alpha}, we observe five different values of $\alpha$ with the corresponding values of $t_s$, and the values of $a(t)$, $\beta(t)$, $\dot{a}(t)$ and $\dot{\beta}(t)$ evaluated in $t = t_s$. The analysis of Table~\ref{Tabela do parametro alpha} shows that the expansion rate $\dot{a}(t)$ strongly depends on the parameter $\alpha$. For lower values of $\alpha$, $\dot{a}(t)$ reaches significantly higher values, indicating a faster expansion of the universe. In contrast, for larger values of $\alpha$, the expansion rate $\dot{a}(t)$ decreases, and the universe becomes less expansive. Finally, the anisotropy parameter $\beta(t)$ tends to stabilize more quickly for high values of $\alpha$, reaching its asymptotic value in shorter time intervals $t_s$. This result implies a faster isotropization accompanied by a slower expansion of the Universe. At $t = t_s$, $\dot{\beta}(t)$ becomes practically zero.
    
    %Tabela
    \begin{table}[h!]
    	\centering
    	\caption{Variation of $t_s$, $a(t_s)$, $\dot{a}(t_s)$, $\beta(t_s)$ and $\dot{\beta}(t_s)$ for different values of $\alpha$, and the fixed values of $a_0=\beta_0=\dot{a}_0=1$, $A=C_r=1$ and $C=5$.}
    	\begin{tabular}{lccccc}
    		\toprule
    		$\alpha$&$t_s$&$a(t_s)\times10^5$&$\dot{a}(t_s)\times10^5$&$\beta(t_s)$&$\dot{\beta}(t_s)\times10^{-13}$\\
    		\midrule
    		\textbf{1.0}&$22.76$&$4.879240$&$2.817031$&$2.678851$&$-4.993200$\\
    		\textbf{0.7}&$22.80$&$5.058214$&$2.920361$&$2.612489$&$-4.964969$\\
    		\textbf{0.5}&$22.84$&$5.266334$&$3.040519$&$2.535702$&$-4.945883$\\
    		\textbf{0.3}&$22.90$&$5.632012$&$3.251643$&$2.401414$&$-4.945998$\\
    		\textbf{0.1}&$23.04$&$6.546819$&$3.779808$&$2.099765$&$-4.949127$\\
    		\botrule
    	\end{tabular}
    	\label{Tabela do parametro alpha}
    \end{table}
	
	%%%%%%%%%%%%%%%%%%%%%%%%%%%%%%%%%%%%%%%%%%%%
	\subsection{Studying the influence of $C_r$}\label{Sec4.2}
\begin{figure}[h]
		\centering
		\includegraphics[width=0.5\textwidth]{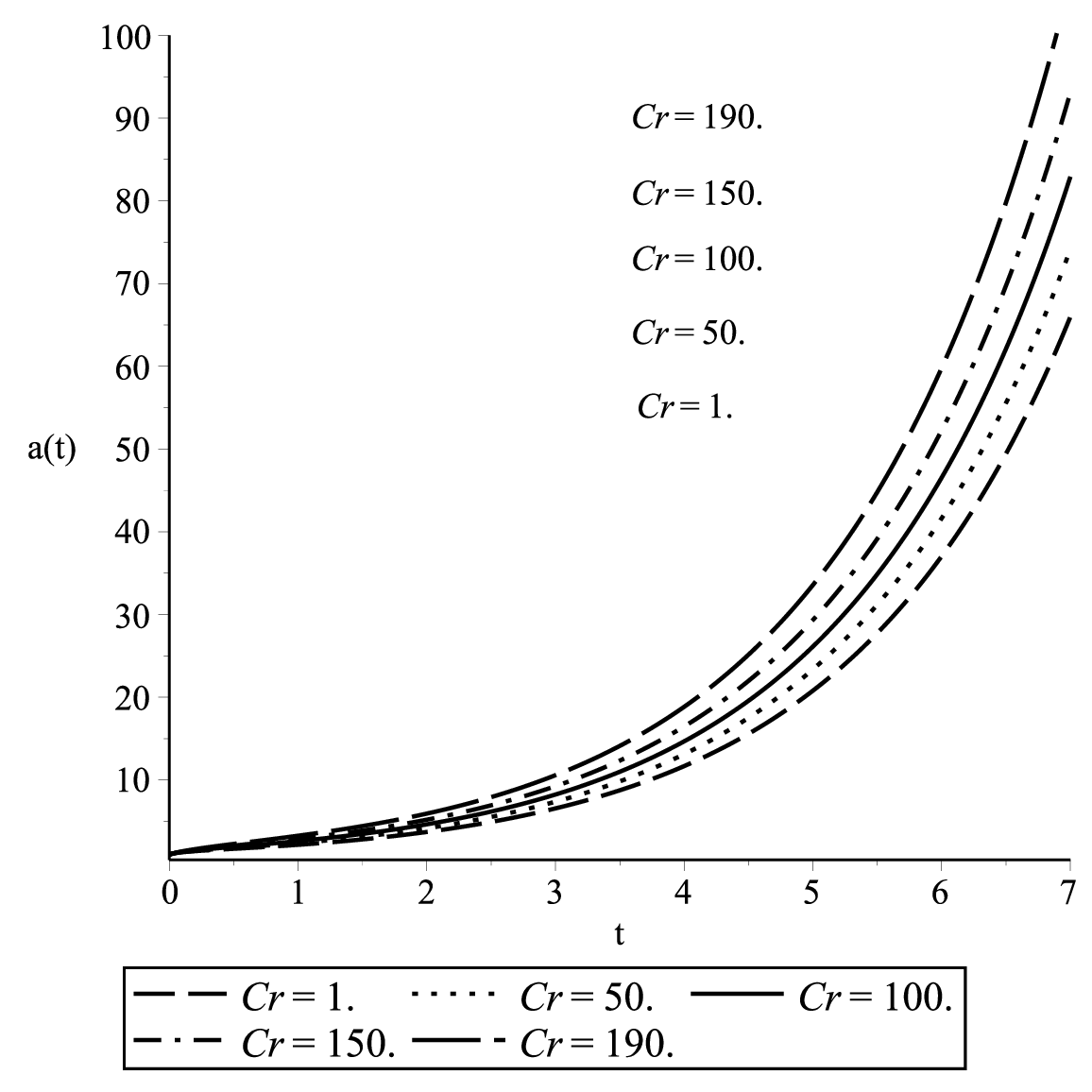}
		\caption{ $a(t)$ as a function of $t$ for different values of $C_r$ com $C=A=1, \alpha=0.7$ and $a_0=\beta_0=1,\dot{a}_0=5$.}
		\label{fig_a. Variando Cr}
	\end{figure}

\begin{figure}[h]
		\centering
		\includegraphics[width=0.5\textwidth]{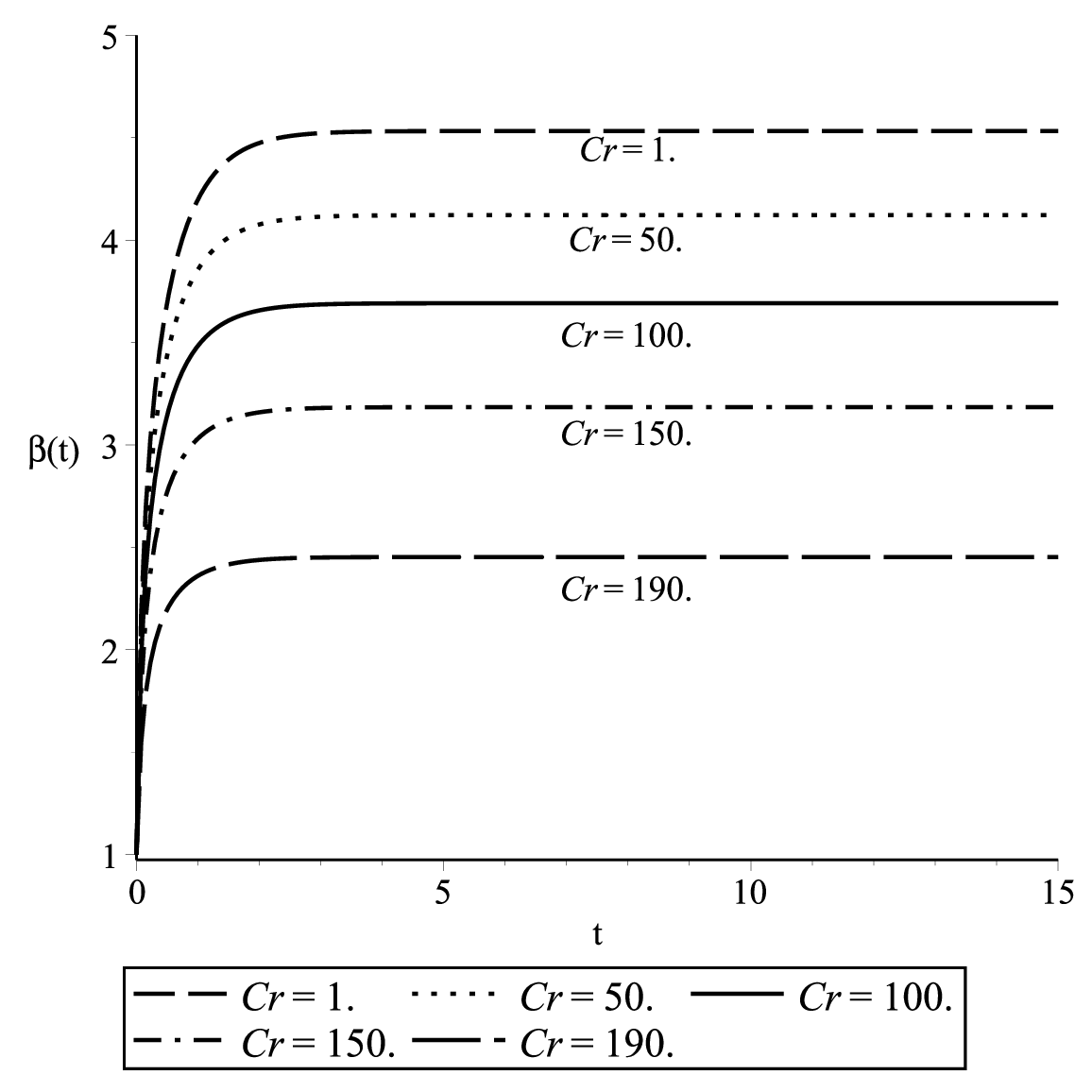}
		\caption{ $\beta(t)$ as a function of $t$ for different values of $C_r$ com $C=A=1, \alpha=0.7$ and $a_0=\beta_0=1,\dot{a}_0=5$.}
		\label{fig_b. Variando Cr}
	\end{figure}

	%%%%%%%%%%%%%%%%%%%%%%%%%%%%%%%%%%%%%%%%%%%%%
	Consider now how variations in the radiation energy density parameter $C_r$ influence the dynamic evolution of the scale factor $a(t)$ and the anisotropy parameter $\beta(t)$. To do that, we must solve the system of equations \eqref{G_00} and \eqref{5Eq importante}. By varying the value of the parameter $C_r$, while keeping the parameters $C$, $A$, and $\alpha$ fixed, we observe the following results: the scale factor $a(t)$ expands more rapidly for larger values of $C_r$, as illustrated in Fig~\ref{fig_a. Variando Cr}. This behavior indicates that the increase in $C_r$ is associated with a higher energy density, which intensifies the expansion of the Universe. On the other hand, the anisotropy parameter $\beta(t)$ converges asymptotically to a positive constant value, whose value depends on $C_r$. For larger values of $C_r$, $\beta(t)$ asymptotically tends towards smaller values; while, for smaller values, $\beta(t)$ stabilizes at relatively larger values, as shown in Fig.~\ref{fig_b. Variando Cr}. The behavior of the system is corroborated by the numerical data presented in Table~\ref{Tabela do parametro Cr}. In it, we present five different values of $C_r$ and the corresponding times $t_s$, defined as the instants when the Universe tends towards isotropization. The table also displays the values of $a(t)$, $\beta(t)$, $\dot{a}(t)$ and $\dot{\beta}(t)$, evaluated at $t = t_s$. The analysis of Table~\ref{Tabela do parametro Cr} shows that the expansion rate $\dot{a}(t)$ strongly depends on the values of the parameter $C_r$. For larger values of $C_r$, $\dot{a}(t)$ reaches significantly higher values, indicating a faster expansion of the Universe. In contrast, for smaller values of $C_r$, the expansion rate $\dot{a}(t)$ is relatively smaller, suggesting a slower expansion. Finally, the anisotropy parameter $\beta(t)$ tends to stabilize more quickly for smaller values of $C_r$, reaching its asymptotic value in a shorter time interval $t_s$. This behavior indicates a faster isotropization accompanied by a slower expansion of the Universe. At $t = t_s$, $\dot{\beta}(t)$ becomes practically zero.
	
	%Tabela
	\begin{table}[h]
		\centering
		\caption{Variation of $t_s$, $a(t_s)$, $\dot{a}(t_s)$, $\beta(t_s)$ and $\dot{\beta}(t_s)$ for different values of $C_r$, and the fixed values of $a_0=\beta_0=1,\dot{a}_0=5$, $A=C=1$ and $\alpha=0.7$.}
		\begin{tabular}{lccccc}
			\toprule
			$Cr$&$t_s$&$a(t_s)\times10^5$&$\dot{a}(t_s)\times10^5$&$\beta(t_s)$&$\dot{\beta}(t_s)\times10^{-13}$\\
			\midrule
			\textbf{1}&$20.90$&$2.017012$&$1.164523$&$4.532473$&$-4.571737$\\
			\textbf{50}&$21.00$&$2.401518$&$1.386517$&$4.121816$&$-4.865088$\\
			\textbf{100}&$21.20$&$3.013628$&$1.739919$&$3.691567$&$-4.752205$\\
			\textbf{150}&$21.40$&$3.797970$&$2.192759$&$3.184131$&$-4.971121$\\
			\textbf{190}&$21.80$&$5.474411$&$3.160653$&$2.451888$&$-4.976992$\\
			\botrule
		\end{tabular}
		\label{Tabela do parametro Cr}
	\end{table}
	
	%%%%%%%%%%%%%%%%%%%%%%%%%%%%%
	\subsection{Studying the influence of $A$}\label{Sec4.3}
\begin{figure}[h!]
		\centering
		\includegraphics[width=0.7\textwidth]{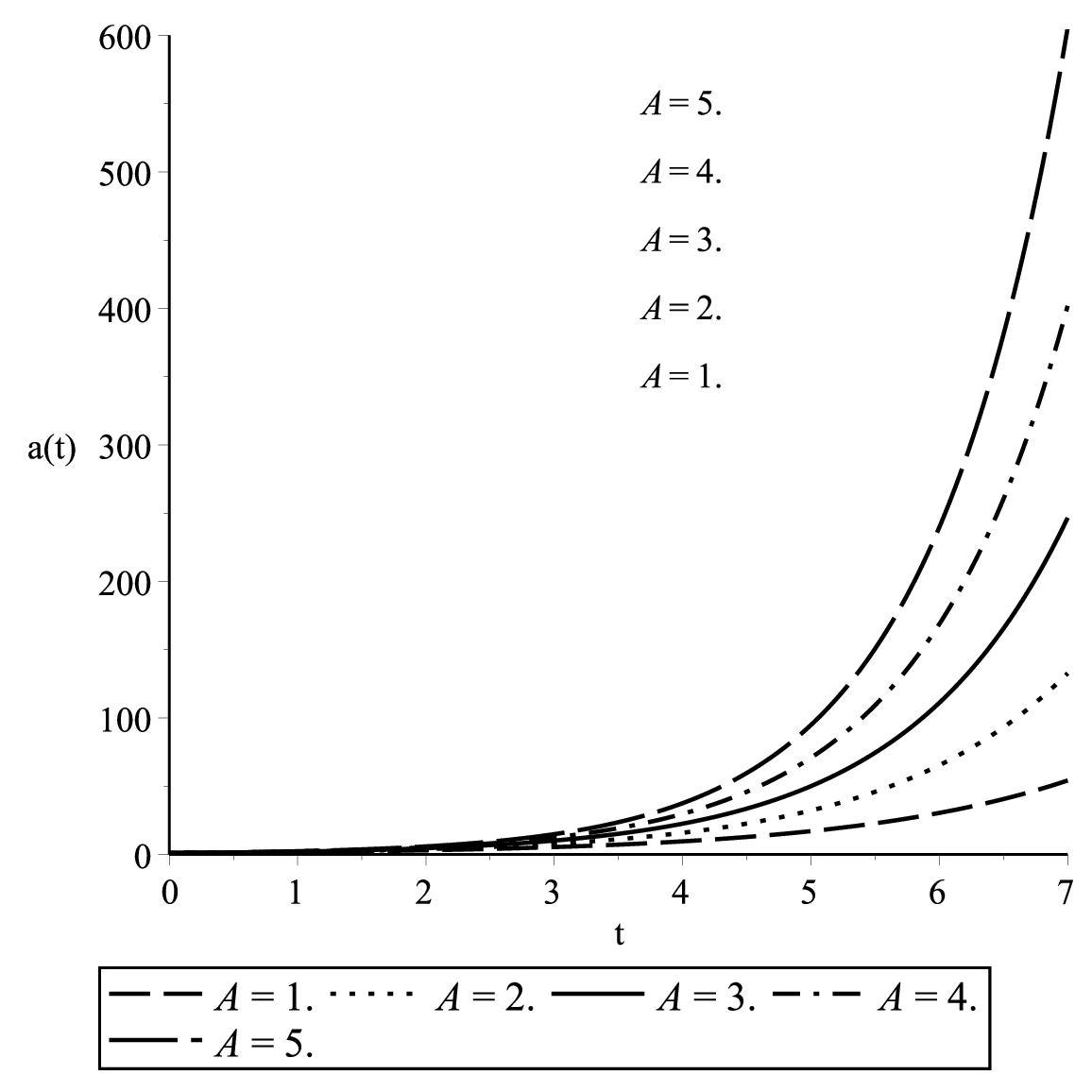}
		\caption{$a(t)$ as a function of $t$ for different values of $A$ with $C=C_r=1, \alpha=0.7$ and $a_0=\beta_0=\dot{a}_0=1$.}
		\label{fig_a. Variando A}
	\end{figure}

\begin{figure}[h!]
		\centering
		\includegraphics[width=0.7\textwidth]{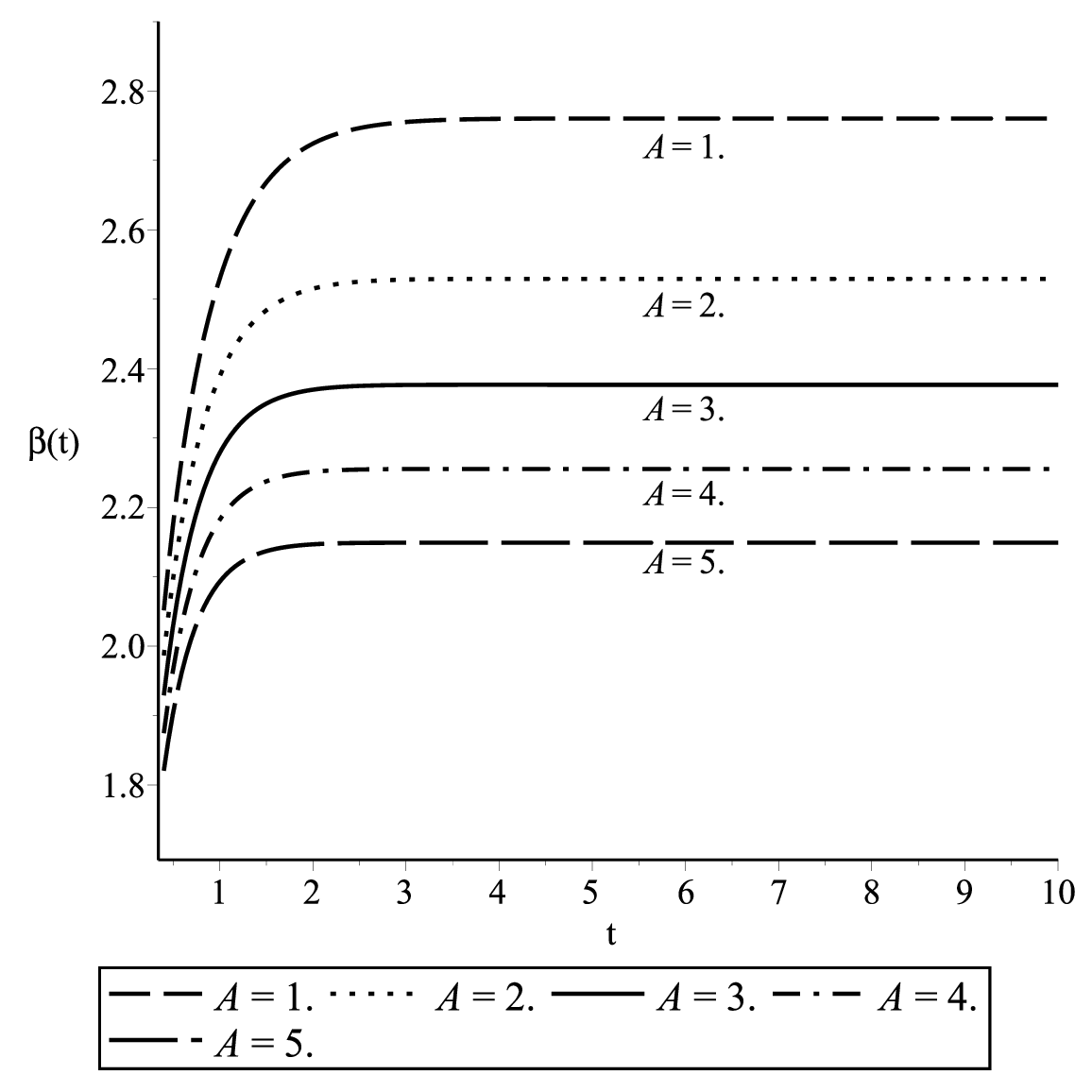}
		\caption{$\beta(t)$ as a function of $t$ for different values of $A$ with $C=C_r=1, \alpha=0.7$ and $a_0=\beta_0=\dot{a}_0=1$.}
		\label{fig_b. Variando A}
	\end{figure}

	%%%%%%%%%%%%%%%%%%%%%%%%%%%%%%
	In this subsection, we investigate how variations in the parameter $A$ influence the dynamic evolution of $a(t)$ and $\beta(t)$. After solving the system of equations \eqref{G_00} and \eqref{5Eq importante}, we realize that by varying the value of $A$, while keeping the parameters $C$, $\alpha$, and $C_r$ fixed, the scale factor $a(t)$ expands more rapidly for higher values of $A$, as illustrated in Fig~\ref{fig_a. Variando A}. This behavior is relevant because, in this regime, the generalized Chaplygin gas begins to behave like a dark energy. On the other hand, the anisotropy parameter $\beta(t)$ converges asymptotically to a positive constant value, whose value strongly depends on the free parameter $A$. For lower values of $A$, $\beta(t)$ tends asymptotically to higher values. However, for higher values of $A$, $\beta(t)$ stabilizes at relatively lower values, as illustrated in Fig.~\ref{fig_b. Variando A}. The behavior of the system is corroborated by the numerical results presented in Table~\ref{Tabela do parametro A}.
    In it, we present five different values of $A$ and the corresponding times $t_s$, defined as the instants when the Universe begins to isotropize. 
	The table also displays the values of $a(t)$, $\beta(t)$, $\dot{a}(t)$ and $\dot{\beta}(t)$, evaluated at $t = t_s$. The analysis of the numerical results indicates that the expansion rate $\dot{a}(t)$ strongly depends on the values of the parameter $A$. For higher values of $A$, $\dot{a}(t)$ reaches significantly higher values, indicating an accelerated expansion of the Universe. In contrast, for lower values of $A$, the expansion rate $\dot{a}(t)$ is relatively smaller, suggesting a slower expansion. Finally, the anisotropy parameter $\beta(t)$ tends to stabilize more quickly for higher values of $A$, reaching its asymptotic value in a shorter time interval $t_s$. This behavior strongly indicates a faster isotropization accompanied by an accelerated expansion of the Universe. At $t = t_s$, $\dot{\beta}(t)$ becomes practically zero.
	
	%Tabela
	\begin{table}[h!]
		\centering
		\caption{Variation of $t_s$, $a(t_s)$, $\dot{a}(t_s)$, $\beta(t_s)$ and $\dot{\beta}(t_s)$ different values of $A$, and the fixed values of $a_0=\beta_0=\dot{a}_0=1$, $C=C_r=1$ and $\alpha=0.7$.}
		\begin{tabular}{lccccc}
			\toprule
			A&$t_s$&$a(t_s)\times10^5$&$\dot{a}(t_s)\times10^5$&$\beta(t_s)$&$\dot{\beta}(t_s)\times10^{-13}$\\
			\midrule
			\textbf{1}&23.0&$5.565985$&$3.213523$&$2.760326$&$-3.536852$\\
			\textbf{2}&$19.0$&$6.488482$&$4.593256$&$2.529290$&$-2.674346$\\
			\textbf{3}&$17.0$&$7.172797$&$5.720812$&$2.376682$&$-2.262650$\\
			\textbf{4}&$15.5$&$6.430743$&$5.581835$&$2.255303$&$-2.920325$\\
			\textbf{5}&$14.5$&$6.325284$&$5.862716$&$2.149106$&$-3.143488$\\
			\botrule
		\end{tabular}
		\label{Tabela do parametro A}
	\end{table}
	
	%%%%%%%%%%%%%%%%%%%%%%%%%%%%%%
	\subsection{Studying the influence of $C$}\label{Sec4.4}
\begin{figure}[h!]
		\centering
		\includegraphics[width=0.7\textwidth]{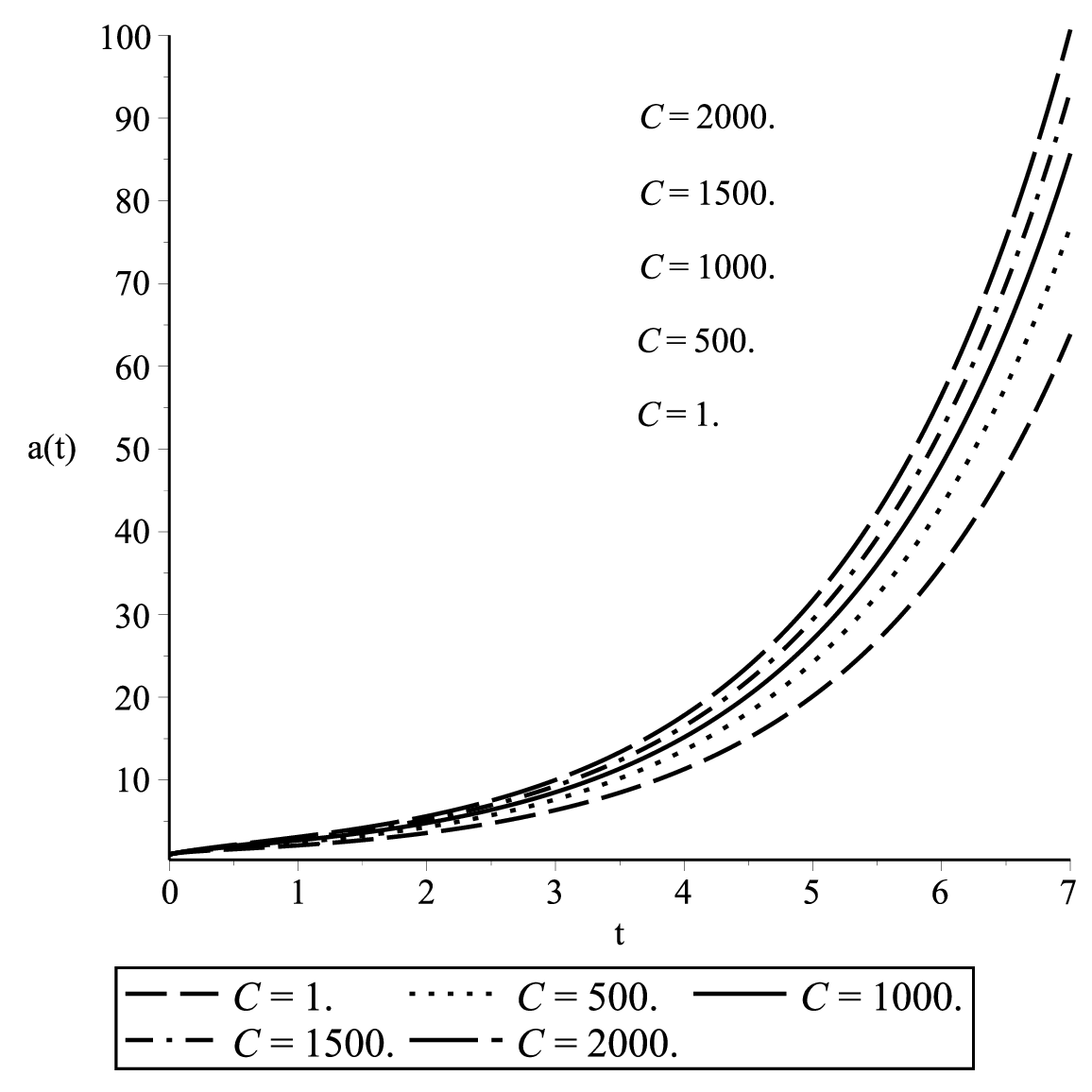}
		\caption{$a(t)$ as a function of $t$ for different values of $C$ with $A=C_r=1, \alpha=0.7$ and $a_0=\beta_0=\dot{a}_0=1$.}
		\label{fig_a. Variando C}
	\end{figure}

\begin{figure}[h!]
		\centering
		\includegraphics[width=0.7\textwidth]{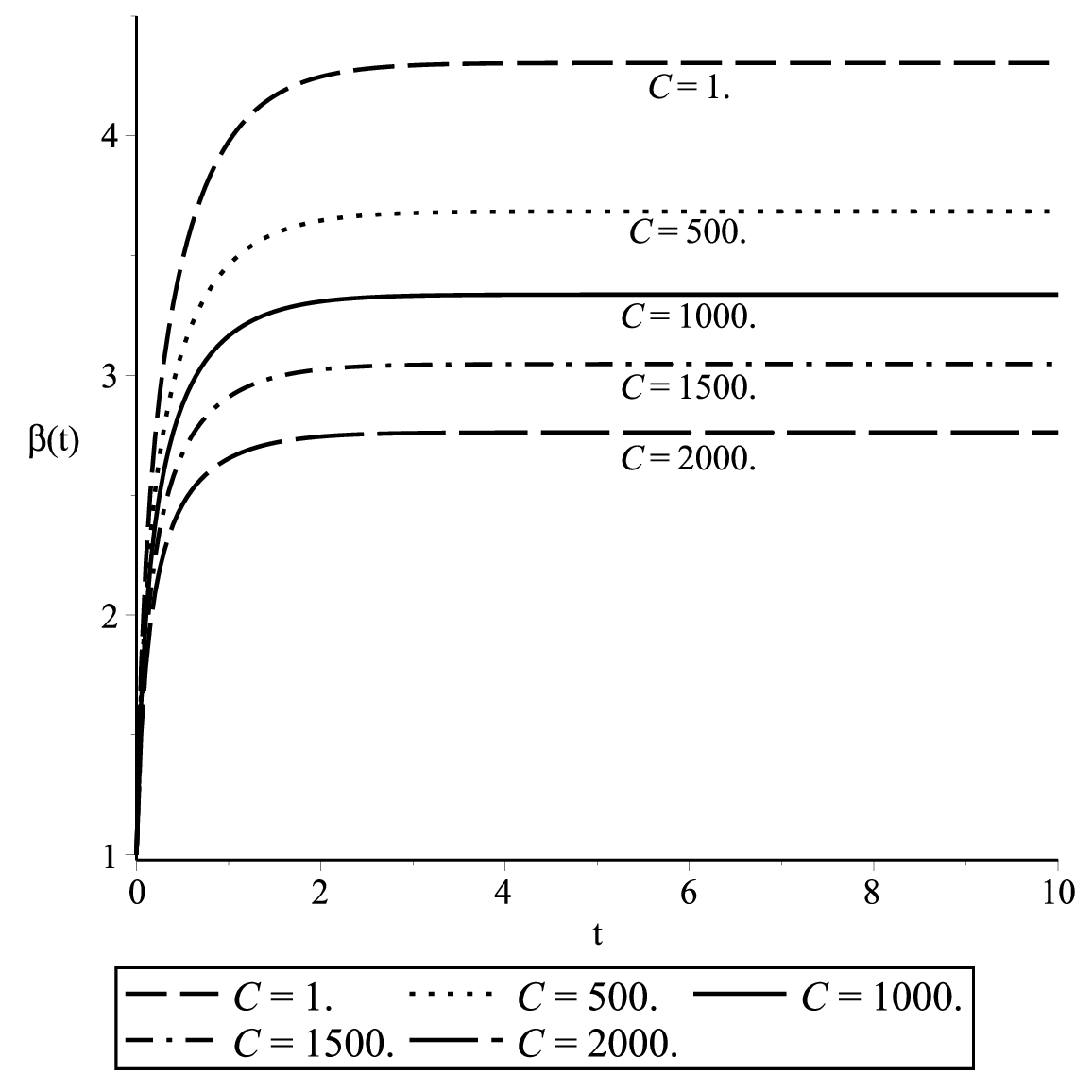}
		\caption{$\beta(t)$ as a function of $t$ for different values of $C$ with $A=C_r=1, \alpha=0.7$ and $a_0=\beta_0=\dot{a}_0=1$.}
		\label{fig_b. Variando C}
	\end{figure}

	%%%%%%%%%%%%%%%%%%%%%%%%%%%%%%
	Finally, we analyze how the parameter $C$, associated with the generalized Chaplygin gas, influences the dynamic evolution of the scale factor $a(t)$ and the anisotropy parameter $\beta(t)$. To do this, we must solve the system of equations \eqref{G_00} and \eqref{5Eq importante}. Changing the value of $C$, while keeping the parameters $C_r$, $\alpha$, and $A$ fixed, we find that the scale factor $a(t)$ expands more rapidly for higher values of $C$, as illustrated in Fig.~\ref{fig_a. Variando C}. On the other hand, the anisotropy parameter $\beta(t)$ converges asymptotically to a positive constant value, whose value strongly depends on the free parameter $C$. For higher values of $C$, $\beta(t)$ tends asymptotically to lower values. However, for lower values of $C$, the anisotropy parameter $\beta(t)$ stabilizes at relatively higher values, as illustrated in Fig.~\ref{fig_b. Variando C}. The behavior of the system is corroborated by the numerical results presented in Table~\ref{Tabela do parametro C}. In it, we present five different values of $C$ and the corresponding times $t_s$, defined as the instants when the Universe begins the isotropization process. The table also displays the values of $a(t)$, $\beta(t)$, $\dot{a}(t)$ and $\dot{\beta}(t)$, evaluated at $t = t_s$. The analysis of the numerical results indicates that the expansion rate $\dot{a}(t)$ strongly depends on the free parameter $C$. For higher values of $C$, $\dot{a}(t)$ reaches higher values, suggesting an accelerated expansion of the Universe. In contrast, for lower values of $C$, the expansion rate $\dot{a}(t)$ is more moderate, characterizing a less accelerated expansion. Finally, the anisotropy parameter $\beta(t)$ tends to stabilize more rapidly for lower values of $C$, reaching its asymptotic value in a shorter time interval $t_s$. This behavior strongly indicates a faster isotropization accompanied by a less accelerated expansion of the Universe. At $t = t_s$, $\dot{\beta}(t)$ becomes practically zero.
	
	Furthermore, it is important to highlight the physical meaning of the free parameter $C$, associated with the generalized Chaplygin gas. From Eq. ~\eqref{Densidade total}, it is clearly observed that, in the early stages of cosmic evolution, when the size of the Universe $a(t)$ was small, the term containing $C$ associated with the Chaplygin gas contributes more significantly than the term containing $A$, to the energy density of GCG. As the Universe expands, this contribution gradually decreases, becoming negligible in later regimes. The term containing $C$ in the expression for the energy density of GCG, when $\alpha=1$ (Chaplygin gas), behaves like the density of a dust fluid.

	%Tabela
	\begin{table}[h!]
		\centering
		\caption{Variation of $t_s$, $a(t_s)$, $\dot{a}(t_s)$, $\beta(t_s)$ and $\dot{\beta}(t_s)$ for different values of $C$, and the fixed values of $a_0=\beta_0=\dot{a}_0=1$, $A=C_r=1$ and $\alpha=0.7$.}
		\begin{tabular}{lccccc}
			\toprule
			$C$&$t_s$&$a(t_s)\times10^5$&$\dot{a}(t_s)\times10^5$&$\beta(t_s)$&$\dot{\beta}(t_s)\times10^{-13}$\\
			\midrule
			\textbf{1}&$21.15$&$2.255666$&$1.302310$&$4.304261$&$-4.594814$\\
			\textbf{500}&$21.30$&$2.960641$&$1.709327$&$3.684296$&$-4.960101$\\
			\textbf{1000}&$21.45$&$3.599073$&$2.077926$&$3.337360$&$-4.749154$\\
			\textbf{1500}&$21.60$&$4.273364$&$2.467228$&$3.047383$&$-4.502365$\\
			\textbf{2000}&$21.75$&$5.027594$&$2.902683$&$2.761819$&$-4.328130$\\
			\botrule
		\end{tabular}
		\label{Tabela do parametro C}
	\end{table}

    %%%%%%%%%%%%%%%%%%%%%%%%%%%%%%%%%%%%%%%%%%%%%%%%%%%%%%%%%%%%%%%%%%%%%%%%
	\section{Dynamical evolution of the model with observational parameters}
    \label{Sec5}
\begin{figure}[h]
		\centering
		\begin{minipage}{0.49\textwidth}
			\centering
			\includegraphics[width=0.7\textwidth]{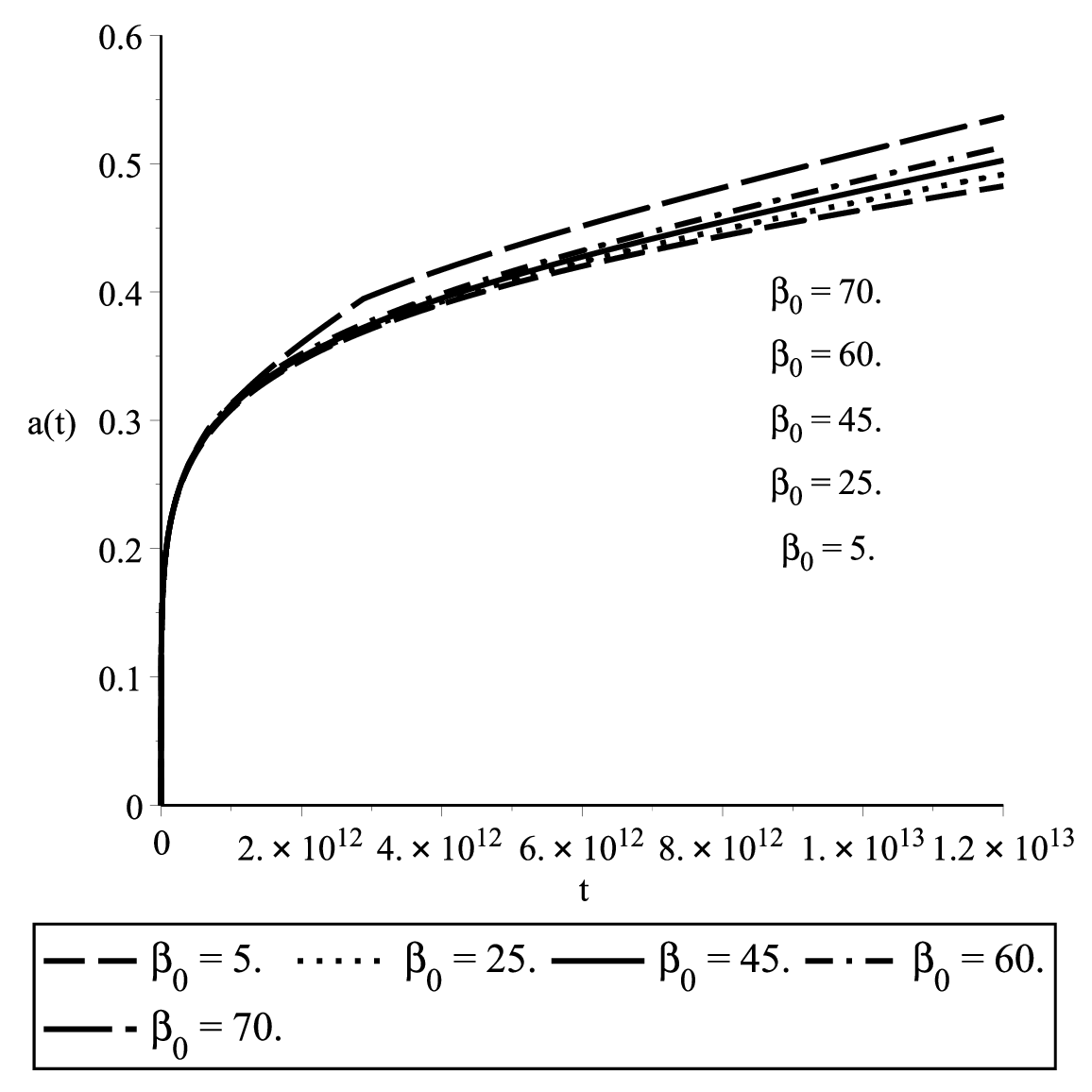}
			\caption{Evolution of the $a(t)$ as a function of the cosmic time $t$, with with $a_0=3.1\times 10^{-7}$ and $\dot{a}_{0}=5\times 10^{16}$. $t$ is given in seconds.}
			\label{fig_a com B_0}
		\end{minipage}
		\hfill
		\begin{minipage}{0.49\textwidth}
			\centering
			\includegraphics[width=0.7\textwidth]{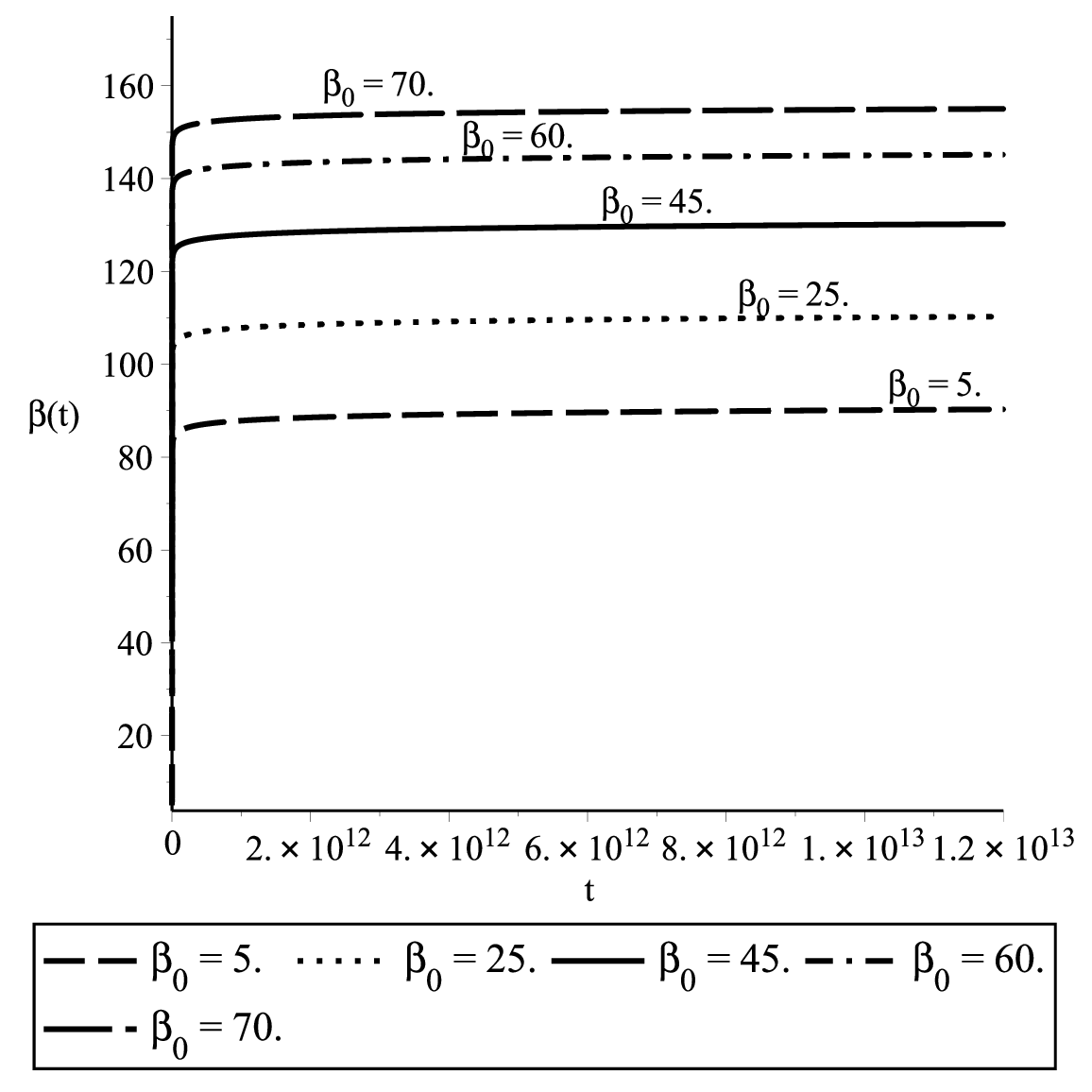}
			\caption{Evolution of the $\beta(t)$ as a function of the cosmic time $t$, with $a_0=3.1\times 10^{-7}$ and $\dot{a}_{0}=5\times 10^{16}$. $t$ is given in seconds.}
			\label{fig_b com B_0}
		\end{minipage}
	\end{figure}

    %%%%%%%%%%%%%%%%%%%%%%%%%%%%%%%%%%%%%%%%%%%%%%%%%%%%%%%%%%%%%%%%%%%%%%%%
    In this section, we investigate the dynamics of the anisotropic KS model in the presence of radiation and a generalized Chaplygin gas, adopting parameter values consistent with observational constraints reported in the literature. The analysis is based on the time evolution of the scale factor $a(t)$ and the anisotropy parameter $\beta(t)$, considering two stages of the model evolution: \textbf{(\textit{i})} the recombination epoch, estimated at approximately $t_{\mathrm{rec}}\equiv3.8\times10^{5}$ years, or $1.2\times10^{13}\mathrm{s}$, associated with the decoupling of photons from baryonic matter and the subsequent free propagation of the cosmic microwave background radiation (CMB); and \textbf{(\textit{ii})} the late-time regime, characterized by the isotropization and the accelerated expansion of the model. The solutions are obtained through the numerical integration of Eqs.~\eqref{Primeira Eq.Dinámica Normalizado} and \eqref{Segunda Eq. Dinámica Normalizado}, now written in SI units. See Appendix \ref{MKS} for more details. In this context, we adopt the observationally constrained values $A_{\mathrm{obs}}=0.759$, $C_{\mathrm{obs}}=0.241$, and $\alpha_{\mathrm{obs}}=0.080$. In the normalized GCG parametrization adopted here, the first two quantities are identified with $B_s=0.759$ and $1-B_s=0.241$, respectively. These values are consistent with observational analyzes of the GCG model based on CMB, Type Ia supernovae, BAO, and observational Hubble data \cite{lu2009observational,liang2011constraints,xu2012revisiting,li2018constraint}. In particular, Li et al.~\cite{li2018constraint} obtained $B_s=0.759$ and $\alpha\simeq0.0801$ from a joint analysis that included Planck 2015 data. We also adopt $C_r=7\times10^{-31} kg/m^3$ for the radiation contribution \cite{hentschke,panpanich}. Keeping these parameters fixed, we observe that during the first stage, the scale factor starts its evolution close to the initial singular regime and subsequently undergoes a period of relatively slow expansion, as illustrated in Fig.~\ref{fig_a com B_0}, followed by a gradual increase in its expansion rate. Within the same regime, the anisotropy parameter $\beta(t)$ evolves towards an asymptotic regime, approaching a positive constant value that depends on the initial condition $\beta_0$. Higher values of $\beta_0$ lead to larger asymptotic values of $\beta(t)$, while smaller values of $\beta_0$ result in smaller asymptotic values, as illustrated in Fig.~\ref{fig_b com B_0}. The evolution of the system is further characterized by the numerical results presented in Table~\ref{tab:resultados_observacionais}. Different values of the initial condition $\beta_0$ are considered and the quantities $a(t)$, $\beta(t)$, $\dot{a}(t)$, and $\dot{\beta}(t)$ are evaluated in the recombination epoch, $t=t_{\mathrm{rec}}$. We observe that, at this stage, $\dot{\beta}(t_{\mathrm{rec}})$ takes small values and decreases as the initial value $\beta_0$ increases. In contrast, $\dot{a}(t_{\mathrm{rec}})$ increases with increasing $\beta_0$, indicating a larger expansion rate for larger values of the initial anisotropy parameter.
    
\begin{figure}[h]
		\centering
		\begin{minipage}{0.49\textwidth}
			\centering
			\includegraphics[width=0.7\textwidth]{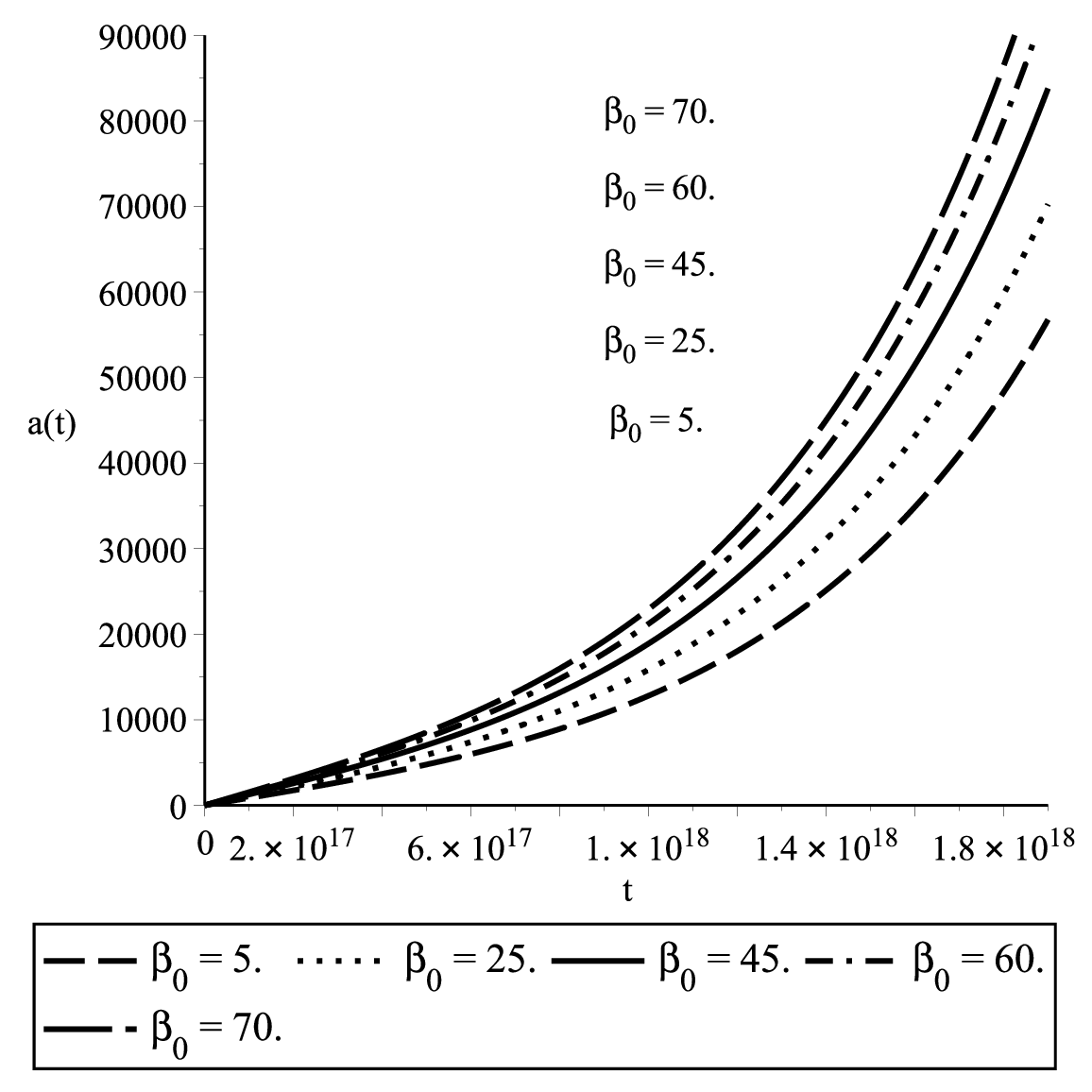}
			\caption{Evolution of the $a(t)$ as a function of the cosmic time $t$, with with $a_0=3.1\times 10^{-7}$ and $\dot{a}_{0}=5\times 10^{16}$. $t$ is given in seconds.}
			\label{fig_a com B_1}
		\end{minipage}
		\hfill
		\begin{minipage}{0.49\textwidth}
			\centering
			\includegraphics[width=0.7\textwidth]{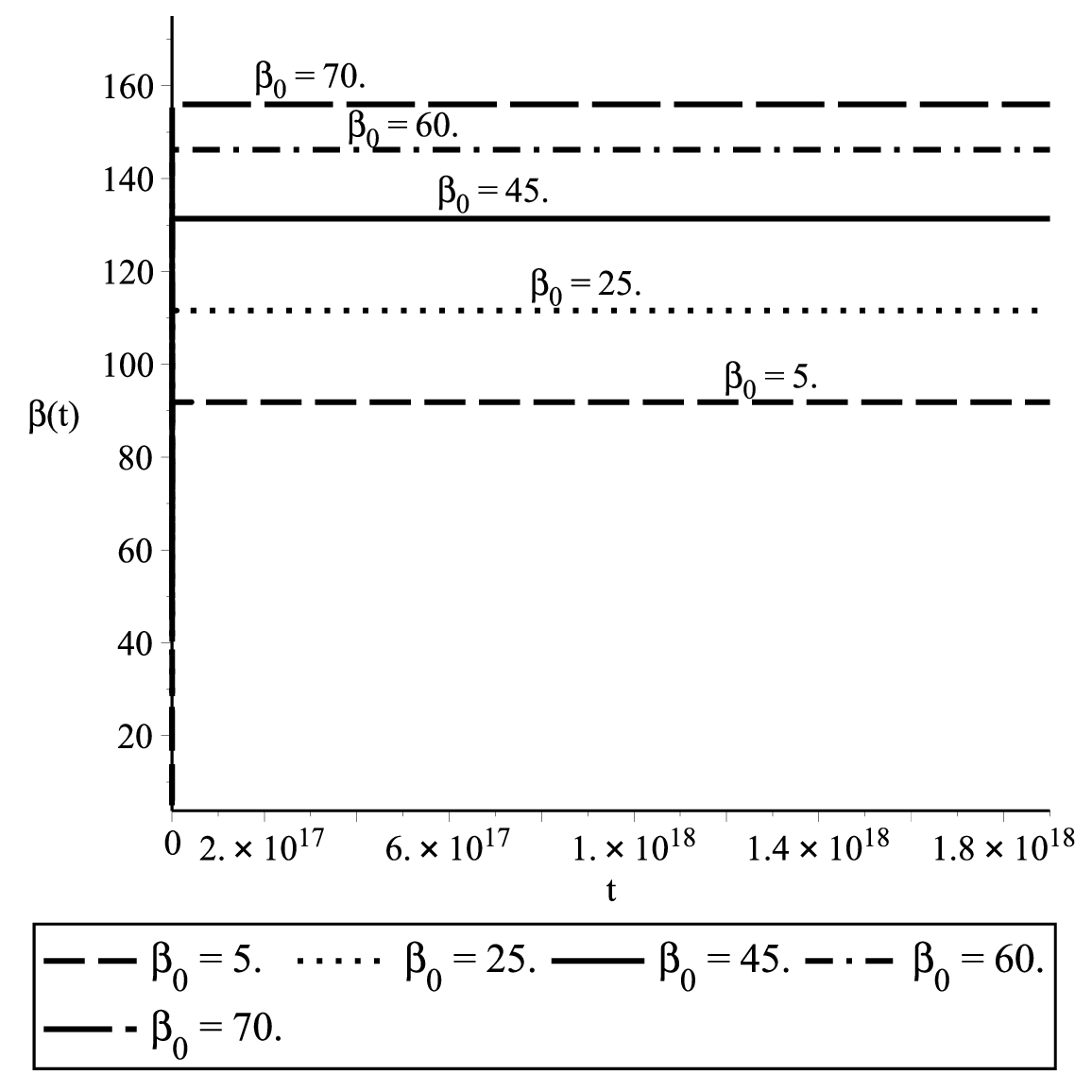}
			\caption{Evolution of the $\beta(t)$ as a function of the cosmic time $t$, with $a_0=3.1\times 10^{-7}$ and $\dot{a}_{0}=5\times 10^{16}$. $t$ is given in seconds.}
			\label{fig_b com B_1}
		\end{minipage}
	\end{figure}

    During the second stage of the evolution, the scale factor exhibits a faster growth. In addition, larger values of the initial condition $\beta_0$ lead to larger values of both $a(t)$ and its rate of evolution, while smaller values of $\beta_0$ correspond to comparatively slower expansion, as shown in Fig.~\ref{fig_a com B_1}. In this regime, the anisotropy parameter $\beta(t)$ preserves its asymptotic behavior, tending to a positive constant value that depends on $\beta_0$. Higher values of $\beta_0$ again lead to larger asymptotic values of $\beta(t)$, whereas smaller values result in smaller asymptotic values, as illustrated in Fig.~\ref{fig_b com B_1}. Furthermore, a reduction in the rate of evolution of the anisotropy is observed, characterized by the tendency $\dot{\beta}(t) \rightarrow 0$ at later times. This behavior indicates that the time evolution of the anisotropy is progressively suppressed during the expansion and provides an indication that the model approaches an isotropic regime. The dynamical evolution of the system is further supported by the numerical results presented in Table~\ref{tab:resultados_observacionais}. The table shows different values of the initial condition $\beta_0$, together with the characteristic time $t_s$, when $\dot{\beta}(t)$ assume values of the order of $10^{-13}$. In our opinion, at these values of $\dot{\beta}(t)$, one may start to consider $\beta(t)$ constant. Therefore, at $t_s$ the universe enters the isotropic regime. 
    The table also presents the corresponding values of $a(t_s)$, $\beta(t_s)$, $\dot{a}(t_s)$, and $\dot{\beta}(t_s)$. For the initial conditions considered, the isotropic regime is reached on time scales of the order of $t_s\sim 10^{12}\mathrm{s}$. 
    This time scale is smaller than the recombination time, $t_{\mathrm{rec}}\simeq3.8\times10^{5}$ years, indicating that, according to the adopted criterion, the anisotropy parameter $\beta(t)$ enters its constant regime before recombination. Therefore, isotropization occurs before the recombination epoch, whereas the accelerated-expansion phase becomes more pronounced only later, as indicated by the evolution of the scale factor shown in Fig.~\ref{fig_a com B_1}.

    Finally, for the adopted set of observational parameters and initial conditions, the KS model with radiation and GCG evolves from an initially anisotropic regime to an asymptotic regime in which $\beta(t)$ approaches a constant value and $\dot{\beta}(t)\rightarrow0$. In somewhat recent times, the contribution of GCG has favored an accelerated expansion of the model. These results are qualitatively consistent with a cosmological evolution characterized by a tendency toward isotropization and accelerated expansion on large scales.

	%segunda parte de figuras: 
	%Tabela
	\begin{table*}[ht!]
		\centering
		\caption{Numerical values of the cosmological quantities evaluated at the characteristic time $t_s$ and at the recombination epoch, $t_{\mathrm{rec}}$, for different initial conditions $\beta_0$. The parameters $C_{\mathrm{obs}}=0.241$, $A_{\mathrm{obs}}=0.759$, $\alpha_{\mathrm{obs}}=0.080$, and $C_r=7\times10^{-31} kg/m^3$ were adopted. $t$ is given in seconds.}
		\label{tab:resultados_observacionais}
		\resizebox{\textwidth}{!}{%
			\begin{tabular}{c c c c c c c c c c}
				\toprule
				& \multicolumn{5}{c}{Asymptotic regime at $t_s$}
				& \multicolumn{4}{c}{Recombination epoch at $t_{\mathrm{rec}}=1.2\times10^{13}\mathrm{s}$} \\
				\cmidrule(lr){2-6}
				\cmidrule(lr){7-10}
				$\beta_0$
				& $t_s\times10^{12}\mathrm{s}$
				& $a(t_s)$
				& $\dot{a}(t_s)\times10^{-14}$
				& $\beta(t_s)$
				& $\dot{\beta}(t_s)\times10^{-13}$
				& $a(t_{\mathrm{rec}})$
				& $\dot{a}(t_{\mathrm{rec}})\times10^{-14}$
				& $\beta(t_{\mathrm{rec}})$
				& $\dot{\beta}(t_{\mathrm{rec}})\times10^{-13}$ \\
				\midrule
				$\mathbf{5}$  & $1.992$ & $0.3463$ & $2.9491$ & $88.5425$  & $4.9990$
				& $0.4824$ & $0.8874$ & $90.2934$  & $0.7705$ \\
				$\mathbf{25}$ & $1.986$ & $0.3472$ & $2.9895$ & $108.5274$ & $4.9987$
				& $0.4915$ & $1.0057$ & $110.2586$ & $0.7414$ \\
				$\mathbf{45}$ & $1.975$ & $0.3484$ & $3.0397$ & $128.5069$ & $4.9996$
				& $0.5025$ & $1.1419$ & $130.2160$ & $0.7090$ \\
				$\mathbf{60}$ & $1.953$ & $0.3502$ & $3.0888$ & $143.4763$ & $4.9994$
				& $0.5129$ & $1.2551$ & $145.1683$ & $0.6829$ \\
				$\mathbf{70}$ & $1.843$ & $0.3534$ & $4.2823$ & $153.4220$ & $4.9976$
				& $0.5364$ & $1.3431$ & $155.0240$ & $0.6423$ \\
				\bottomrule
			\end{tabular}%
		}
	\end{table*}
    
	%%%%%%%%%%%%%%%%%%%%%%%%%%%%%%%%%%%%%%%%%%%%%%%%%%%
	\section{Conclusions}\label{Sec6}
    %%%%%%%%%%%%%%%%%%%%%%%%%%%%%%%%%%%%%%%%%%%%%%%%%%%

    In this work, we studied the dynamical evolution of a homogeneous and anisotropic $KS$ cosmological model, considering general relativity as the gravitational theory, and including two distinct fluids: radiation and generalized Chaplygin gas. Our main objective was to investigate whether the model tends towards a homogeneous and isotropic standard behavior during its evolution. Furthermore, we sought to understand how the model parameters quantitatively influence the isotropization process. To simplify our analysis, we used the Misner parametrization of the $KS$ metric Eq.~\eqref{metric}. In this parametrization, the metric has two relevant metric functions: the scale factor $a(t)$, which describes the expansion of the Universe, and the anisotropy parameter $\beta(t)$, which measures the spatial anisotropy of the model. Isotropization occurs in two possible cases: (\textbf{\textit{i}}) when $\beta(t) \to 0$ or (\textbf{\textit{ii}}) when $\beta(t)$ asymptotically tends to a finite constant value. In these two situations, we are able to recover the standard $FRW$ metric with closed geometry and positive curvature $k=+1$. Initially, we obtained the effective Hamiltonian of the model. From it we derived the three independent Einstein field equations. Before the numerical solution, we identified the qualitative behavior of the solutions from the Hamiltonian, constructing phase portraits of the model. We verified that the solutions are of the expansive type, exhibiting isotropization and accelerated expansion of the Universe after a certain time interval. Next, we numerically solved the system formed by two second-order ordinary differential equations, obtained from the field equations. Our attention focused on the expansive solutions. We investigated how the parameters of the GCG ($A, C, \alpha$) and the parameter associated with the energy density of the radiation fluid ($C_r$) influence the dynamic evolution of the scale factor $a(t)$ and the anisotropy parameter $\beta(t)$.

    As a result, we found that all solutions are expansive 
    with $a(t)$ starting the expansion from a small finite value.
    The Universe expands rapidly until the gravitational force ceases to restrain the growth, and the influence of the GCG leads to an accelerated expansion of the Universe, promoting a high degree of isotropy. The anisotropy parameter $\beta(t)$ also starts from a small, finite value, increases rapidly, and after some time asymptotically tends to a positive constant value. This behavior ensures that the solutions are asymptotically isotropic, indicating that the metric $KS$ Eq.~\eqref{metric} naturally evolves into a standard metric of type $FRW$, with $a(t)$ playing the role of a scale factor. The asymptotic constancy of $\beta(t)$ was confirmed by calculating its time rate $\dot{\beta}$, which decreases rapidly, reaching significantly small values on the order of $10^{-13}$, evidencing the high degree of isotropy achieved by the model. Based on the numerical simulations, we identified the following general behaviors: (\textbf{\textit{i}}) the scale factor $a(t)$ expands more rapidly for higher values of the parameters $A, C$ of GCG and $C_r$ of radiation; (\textbf{\textit{ii}}) $a(t)$ expands more rapidly for lower values of $\alpha$; (\textbf{\textit{iii}}) the anisotropy parameter $\beta(t)$ tends asymptotically to higher values when the value of $\alpha$ increases or when the values of the parameters $C, C_r$ decrease; and (\textbf{\textit{iv}}) $\beta(t)$ tends asymptotically to lower values as the value of $A$ increases. These results indicate that the proposed model can describe an initial anisotropic stage of the Universe, which naturally evolves into an isotropic state at later times due to accelerated expansion induced by the generalized Chaplygin gas. The asymptotic behavior obtained demonstrates that isotropization is an inevitable consequence of cosmological dynamics in this scenario.

    \appendix

    \section{Dynamical Equations in SI Units}
    \label{MKS}
    
    To investigate KS dynamics using parameters consistent with observational constraints, we restore the gravitational factor $8\pi G$ and introduce the densities and cosmological parameters units required to make Eqs.~\eqref{G_00} and \eqref{5Eq importante} dimensionally consistent in SI units. The first dynamical equation then becomes the following.
    \begin{equation}
		\begin{aligned}
			\frac{-\dot{\beta}^2}{4}+\frac{\dot{a}\dot{\beta}}{a}+\frac{3\dot{a}^2}{a^2}&=8\pi G \rho_{GCG}\Bigg[A_{obs}+C_{obs}\bigg(a^3e^{\beta/2}\bigg)^{-1-\alpha_{obs}}\Bigg]^{\frac{1}{1+\alpha_{obs}}}\\
			&\qquad + 8\pi G C_r\frac{e^{-2\beta/3}}{a^4}-8\pi G \Omega_{k}\rho_{c}\frac{e^{-\beta}}{a^2}.
		\end{aligned}
		\label{Primeira Eq.Dinámica Normalizado}
	\end{equation}
    Similarly, the equation governing the evolution of the anisotropy parameter is written as
    \begin{equation}
        a^2\ddot{\beta}+\frac{a^2\dot{\beta}^2}{2}+3a\dot{a}\dot{\beta}+8\pi G \Omega_{k}\rho_{c}e^{-\beta}=0.
		\label{Segunda Eq. Dinámica Normalizado}
    \end{equation}
    Here, $G$ is Newton's gravitational constant, $\rho_{\mathrm{GCG}}$ is the current GCG density, and $\rho_{\mathrm{c}}$ is the critical density. The parameter $\Omega_k$ is the current curvature density parameter, while $A_{\mathrm{obs}}$, $C_{\mathrm{obs}}$, and $\alpha_{\mathrm{obs}}$ denote the current GCG parameters adopted from observational constraints. The quantity $C_r$ is the current radiation density. The numerical values used in the integrations are summarized in Table~\ref{Tabela_de_Valores}
\cite{lu2009observational,liang2011constraints,xu2012revisiting,li2018constraint,hentschke,panpanich}.
    \begin{table*}[ht!]
		\centering
		\caption{Cosmological parameters and physical constants adopted in the numerical integration.}
		\label{Tabela_de_Valores}
		\resizebox{\textwidth}{!}{%
			\begin{tabular}{c c c c c c}
				\toprule
				\multicolumn{3}{c}{GCG parameters}
				&
				\multicolumn{3}{c}{Cosmological quantities and physical constant}\\
				\cmidrule(lr){1-3}  \cmidrule(lr){4-6}
				Symbol & Adopted value & SI unit & Symbol & Adopted value & SI unit\\
				\midrule
				$A_{\mathrm{obs}}$ & $0.759$ & -- & $C_r$ & $7.0\times10^{-31}$ & $\mathrm{kg\,m^{-3}}$ \\
				$C_{\mathrm{obs}}$ & $0.241$ & -- & $\Omega_k$ & $1.0\times10^{-3}$ & --\\
				$\alpha_{\mathrm{obs}}$ & $0.080$ & -- & $\rho_{\mathrm{c}}$ & $8.5\times10^{-27}$ & $\mathrm{kg\,m^{-3}}$\\
				$\rho_{\mathrm{GCG}}$ & $6.0\times10^{-27}$ & $\mathrm{kg\,m^{-3}}$ & $G$ & $6.67430\times10^{-11}$ & $\mathrm{m^{3}\,kg^{-1}\,s^{-2}}$\\
				\bottomrule
			\end{tabular}%
		}
	\end{table*}
	%----------------------------------------------------------------------
	\backmatter    % Parte final (sem numeração de capítulos)
	
	\bmhead{Supplementary information}% Para escrever uma informacao extra. Escrever...
	
	\section*{Declarations}

    {\bf Acknowledgments}. R. Hincho Laucata thanks Coordena\c{c}\~{a}o de Aperfei\c{c}oamento de Pessoal de N\'{i}vel Superior (CAPES) and Universidade Federal de Juiz de Fora (UFJF) for his scholarship. G. Oliveira-Neto thanks FAPEMIG (APQ-06640-24) for partial financial support.

    {\bf Corresponding author}. G. Oliveira-Neto.

    {\bf Authors Contributions}. All authors contributed equally to the conception and elaboration of this manuscript.

    {\bf Funding and/or Conflicts of interests/Competing interests}. The authors declare that there is no conflict of interest regarding the publication of this manuscript.

    {\bf Data Availability}. No datasets were generated or analyzed during the current study. 

	{\bf Ethical Approval} Not applicable in our research shown here.

	%%%%===================================================================

\begin{thebibliography}{99}

\bibitem{DinvernoVickers2022} Ray d'Inverno and James Vickers,
{\it Introducing Einstein's Relativity: A Deeper Understanding},
Second Edition, (Oxford University Press, Oxford, 2022).
%chapter      = {12},
%isbn         = {9780198862021}

\bibitem{MTW1973} Charles W. Misner and Kip S. Thorne and John A. Wheeler, {\it Gravitation}, First Edition (W. H. Freeman and Company, San Francisco, 1973).
%pages     = {1279},
%note      = {p.~809}

\bibitem{kester} C. E. Kester, A. Bernui and W. S. Hipólito-Ricaldi, {\it Probing the statistical isotropy of the universe with Planck data of the cosmic microwave background}, Astronomy and Astrophysics {\bf 683}, A176 (2024).
%(doi: 10.1051/0004-6361/202348160)

\bibitem{barrow} J. D. Barrow and H. Kodama, {\it The isotropy of compact universes}, Classical and Quantum Gravity {\bf 18}, 1753–1766 (2001).
%(doi: 10.1088/1361-6382/adfff9).

\bibitem{adhav} K. S. Adhav, {\it Binary mixture of anisotropic dark energy and perfect fluid in Kantowski-Sach universe}, The European Physical Journal Plus {\bf 126}, 103 (2011).
%(doi: 10.1088/1361-6382/adfff9).

\bibitem{simpliciano} S. C. dos Reis and I. L. Shapiro, {\it Cosmic anisotropy with reduced relativistic gas}, The European Physical Journal C {\bf 78}, 145 (2018).
%(doi: 10.1088/1361-6382/adfff9)

\bibitem{gil0} G. Oliveira-Neto and D. L. Canedo, {\it An Anisotropic Kantowski–Sachs Universe with Radiation, Dust and a Phantom Fluid}, Brazilian Journal of Physics {\bf 52}, 130 (2022).
%(doi: 10.1007/s13538-022-01137-0).

\bibitem{parnovsky} S. L. Parnovsky, {\it The Big Bang could be anisotropic. The case of Bianchi I model}, Classical and Quantum Gravity {\bf 40}, 135005 (2023).
%(doi: 10.1088/1361-6382/acd7c2).

\bibitem{thiago} G. Oliveira-Neto and T. M. Abreu, {\it A noncommutative Bianchi I model with radiation}, International Journal of Modern Physics D {\bf 33}, 2450032 (2024).
%(doi: 10.1142/S0218271824500329)

\bibitem{thiago1} T. M. Abreu, G. Oliveira-Neto, A. C. R. Mendes and S. C. Reis, {\it Noncommutative effects in Bianchi I
cosmology with reduced relativistic gas}, Classical and Quantum Gravity {\bf 42}, 185010 (2025).
%(doi: 10.1088/1361-6382/adfff9)

\bibitem{yuri} G. Oliveira-Neto, G. and Y. S. Apaza, {\it Noncommutative Bianchi I and III Cosmology Models: Radiation Era
Dynamics and $\gamma$ Estimation}, Universe {\bf 11}, 408 (2025).
%(doi: 10.3390/universe11120408)

\bibitem{faranas} G. Faranas and A. Vilenkin, {\it Jackiw-Teitelboim and Kantowski-Sachs quantum cosmology}, Journal of Cosmology and Astroparticle Physics {\bf 03}, 056 (2022).

\bibitem{faranas1} G. Faranas and A. Vilenkin, {\it The tunneling wavefunction in Kantowski-Sachs quantum cosmology}, Journal of Cosmology and Astroparticle Physics {\bf 08}, 069 (2022).

\bibitem{hawking} J. B. Hartle and S. W. Hawking, {\it Wave function of the universe}, Physical Review D {\bf 28}, 2960 (1983).

\bibitem{vilenking} A. Vilenkin, {\it Boundary conditions in quantum cosmology}, Physical Review D {\bf 33}, 3560 (1986).

\bibitem{vilenking1} A. Vilenkin, {\it Quantum cosmology and the initial state of the universe},
Physical Review D {\bf 37}, 888 (1988).

\bibitem{vilenking2} A. Vilenkin, {\it Approaches to quantum cosmology}, Physical Review D {\bf 50}, 2581 (1994).

\bibitem{singh} A. Singh, {\it Qualitative aspects of Kantowski–Sachs cosmologies with cosmological constant and stiff fluid}, The European Physical Journal Plus {\bf 138}, 188 (2023).
%(doi: 10.1140/epjp/s13360-023-03829-1)

\bibitem{paliathanasis} A. Paliathanasis, {\it Phase-space analysis of the anisotropies in Weyl integrable spacetime: Kantowski–Sachs cosmology}, Mathematical Methods in the Applied Sciences
{\bf 47}, 4076–4093 (2023).

%\bibitem{luciano} G. G. Luciano, {\it Saez–Ballester gravity in Kantowski–Sachs Universe: A new
%reconstruction paradigm for Barrow Holographic Dark Energy}, Physics of the Dark Universe {\bf 41},
%101237 (2023).

\bibitem{vinutha} T. Vinutha and K. V. Vasavi and K. Niharika and G. Satyanarayana, 
{\it Kantowski–Sachs perfect fluid cosmological model in R2- Gravity}, Indian Journal of Physics
{\bf 97(5)}, 1621–1632 (2023).
%(doi: 10.1007/s12648-022-02470-5).

\bibitem{oliveira} V. G. Oliveira, G. Oliveira-Neto and I. L. Shapiro, {\it Kantowski–Sachs Model with a Running Cosmological Constant and Radiation}, Universe {\bf 10}, 83 (2024).
%(doi: 10.3390/universe10020083).

\bibitem{millano} A. D. Millano, K. Dialektopoulos, N. Dimakis, A. Giacomini, H. Shababi, A. Halder and A. Paliathanasis, {\it Kantowski-Sachs and Bianchi III dynamics in $f(Q)$ gravity},
Physical Review D {\bf 109}, 124044 (2024).
%(doi: 110.1103/PhysRevD.109.124044).

\bibitem{ks} R. Kantowski and R. K. Sachs, {\ it Some Spatially Homogeneous Anisotropic Relativistic Cosmological Models}, Journal of Mathematical Physics {\bf 7}, 443 (1966).

\bibitem{misner} Charles W. Misner, {\it Minisuperspace}, in Magic without magic: John Archibald Wheeler, Editor J. R. Klauder, (W. H. Freeman and Company, San Francisco, 1972), p. 449.
%  isbn         = {9780198862021}

\bibitem{riess0} A. G. Riess et al., {\it Observational evidence from supernovae for an accelerating universe and a cosmological constant}, The Astronomical Journal {\bf 116}, 1009 (1998).

\bibitem{perlmutter} S. Perlmutter et al., {\it Measurements of $\Omega$ and $\Lambda$ from 42 high-redshift supernovae}, The Astrophysical Journal {\bf 517}, 565 (1999).

\bibitem{chaplygin} S. Chaplygin, {\it On Gas Jets}, Scientific Memoirs, Moscow University Mathematical Physics {\bf 21}, 1-121 (1904).

\bibitem{kamenshchik2001alternative} A. Kamenshchik, U. Moschella and V. Pasquier, {\it An alternative to quintessence}, Physics Letters B {\bf 511}, Ns. 2-4, 265-268 (2001).

\bibitem{bilic} N. Bilic, G. B. Tupper and R. D. Viollier, {\it Unification of dark matter and dark energy: the inhomogeneous Chaplygin gas}, Physics Letters B {\bf 535}, 17 (2002).

\bibitem{bento} M. C. Bento, O. Bertolami and A. A. Sen, {\it Generalized Chaplygin gas, accelerated expansion, and dark-energy-matter unification}, Physical Review D {\bf 66}, 043507 (2002).

\bibitem{paulo} M. Bouhmadi-Lopez and P.V. Moniz, {\it FRW quantum cosmology with a generalized Chaplygin gas}, Physical Review D {\bf 71}, 063521 (2005).

\bibitem{gil} G. A. Monerat, G. Oliveira-Neto, E. V. Correa Silva, L. G. Ferreira Filho, P. Romildo Júnior, J. C. Fabris, R. Fracalossi, S. V. B. Goncalves and F. G. Alvarenga, {\it The dynamics of the early universe and the initial conditions for inflation in a model with radiation and a Chaplygin gas}, Physical Review D {\bf 76}, 024017 (2007).

\bibitem{pedram0} P. Pedram and S. Jalalzadeh, {\it Quantum FRW cosmological solutions in the presence of Chaplygin gas and perfect fluid}, Physics Letters B {\bf 659}, 6-13 (2008).

\bibitem{pedram} P. Pedram and S. Jalalzadeh, {\it Chaplygin gas quantum universe in the presence of the cosmological constant}, General Relativity and Gravitation {\bf 42}, 745-762 (2010).

\bibitem{majumder} B. Majumder, {\it Quantum scalar-metric cosmology with Chaplygin gas}, Physics Letters B {\bf 697}, 101-106 (2011).

\bibitem{pedram1} H. Ardehali and P. Pedram, {\it Chaplygin gas Ho\v{r}ava-Lifshitz quantum cosmology}, Physical Review D {\bf 93}, 043532 (2016).

\bibitem{pedram2} H. Shababi and P. Pedram, {\it Ho\v{r}ava–Lifshitz quantum cosmology in the presence of Chaplygin gas: Exact solutions and the late-time acceleration}, International Journal of Modern Physics D {\bf 26}, 1750081 (2017).

\bibitem{monerat} G. A. Monerat, C. G. M. Santos, G. Oliveira-Neto, E. V. Corrêa Silva,
L. G. Ferreira Filho, {\it The dynamics of the early universe in a model with radiation and a generalized Chaplygin gas: Early universe with radiation and a generalized Chaplygin gas}, The European Physical Journal Plus {\bf 136}, 34 (2021).
%(doi: 10.1140/epjp/s13360-020-00996-3).

\bibitem{Wald1984} Robert M. Wald, {\it General Relativity},
(University of Chicago Press, Chicago, 1984).
%pages     = {62},
%isbn      = {978-0-226-87033-5}

\bibitem{Weinberg2008} Steven Weinberg, {\it Cosmology}, (Oxford University Press, Oxford, 2008).
%isbn      = {978-0-19-852682-7},
%pages     = {593}

\bibitem{Sundermeyer1982} Kurt Sundermeyer, {\it Constrained Dynamics: With Applications to Yang-Mills Theory, General Relativity, Classical Spin, Dual String Model}, Lecture Notes in Physics {\bf 169} (Springer-Verlag, Berlin, 1982). 
%pages=318.
%(isbn: 0387119477, 9780387119472),
%(doi: 10.1007/BFb0036225).

\bibitem{lu2009observational} J. Lu, Y. Gui and L. Xu, {\it Observational constraint on generalized Chaplygin gas model}, The European Physical Journal C {\bf 63}, 349-354 (2009).
%(doi: 10.1140/epjc/s10052-009-1118-8).

\bibitem{liang2011constraints} N. Liang, L. Xu and Z. H. Zhu, {\it Constraints on the generalized Chaplygin gas model including gamma-ray bursts via a Markov Chain Monte Carlo approach}, Astronomy and Astrophysics {\bf 527}, A11 (2011).
%(doi: 10.1051/0004-6361/201015919)

\bibitem{xu2012revisiting} L. Xu, J. Lu, and Y. Wang, {\it Revisiting Generalized Chaplygin Gas as a Unified Dark Matter and Dark Energy Model}, The European Physical Journal C \textbf{72}, 1883 (2012).
%doi:10.1140/epjc/s10052-012-1883-7.

\bibitem{li2018constraint} H. Li, W. Yang and Y. Wu, {\it Constraint on the generalized Chaplygin gas as a unified dark fluid model after Planck 2015}, Physics of the Dark Universe {\bf 22}, 60-66 (2018).

\bibitem{hentschke} R. Hentschke and C. H\"{o}lbling, {\it A Short Course in General Relativity and Cosmology}, Undergraduate Lecture Notes in Physics (Springer, Cham, 2020), p. 163. %doi.org/10.1007/978-3-030-46384-7. 

\bibitem{panpanich} S. Panpanich, P. Burikham, S. Ponglertsakul and L. Tannukij, {it Resolving Hubble Tension with Quintom Dark Energy Model}, Chinese Physics C {\bf 45}, No. 1, 015108 (2021). 


%(doi: 10.1016/j.dark.2018.09.001)

%\bibitem{canedo2025resonances} D. L. Canedo, G. Oliveira-Neto, G. A. Monerat, and E. V. Silva, {\it Resonances in the early Universe}, arXiv preprint arXiv:2512.13621 (2025).


%\bibitem{Hilbert1915} D. Hilbert, {\it Nachrichten von der Gesellschaft der Wissenschaften zu Göttingen, Mathematisch-Physikalische Klasse}, Die Grundlagen der Physik . (Erste Mitteilung.) {\bf 1915}, {395-408} (1915), http://eudml.org/doc/58946

%% book, authored

%\bibitem{dinverno1992} Ray d'Inverno, {\it Introducing Einstein's Relativity}, (Oxford University Press, Oxford, 1992).
%207-208.
%(isbn: 978-0-19-859686-3).

%\bibitem{liddle} A. R. Liddle, {\it An Introduction to Modern Cosmology}, First Edition, (Wiley \& Sons, Chichester, 2015).
%  isbn         = {9780198862021}


%PAPER WITH VALUES PARAMETER OBSERVATIONAL.

\end{thebibliography}
\end{document}